\documentclass[fleqn,usenatbib]{mnras}

\usepackage{newtxtext,newtxmath}

\usepackage[T1]{fontenc}
\usepackage{ae,aecompl}

\usepackage{graphicx}	
\usepackage{amsmath}	
\usepackage{amssymb}	
\usepackage{subfig}

\newcommand{\cxo}{{\it Chandra}}
\def\xmm{\textit{XMM-Newton}}
\newcommand{\ergcm}[1]{erg\,cm$^{-2}$\,s$^{-1}$}
\newcommand{\dem}{DEM\,S5}

\def\HI{\hbox{H\,{\sc i}}}
\def\HII{\hbox{H\,{\sc ii}}}
\newcommand{\SII}{[S\,{\sc ii}]}
\newcommand{\OIII}{[O\,{\sc iii}]}
\newcommand{\Halpha}{H${\alpha}$}
\newcommand{\ArII}{[Ar\,{\sc ii}]}
\newcommand{\FeII}{[Fe\,{\sc ii}]}
\newcommand{\D}{$^\circ$}

\usepackage{xcolor}

\title[Pulsar-powered Bow Shock Nebula in SMC \dem]{ Discovery of a Pulsar-powered Bow Shock Nebula in the Small Magellanic Cloud Supernova Remnant \dem}

\author[R. Z. E. Alsaberi et al.]{
Rami Z. E. Alsaberi,$^{1}$\thanks{E-mail: r.alsaberi@westernsydney.edu.au}
C. Maitra,$^{2}$
M. D. Filipovi\'c,$^{1}$
L. M. Bozzetto,$^{1}$
F. Haberl,$^{2}$
\newauthor
P. Maggi,$^{3}$
M. Sasaki,$^{4}$
P. Manjolovi\'c,$^{1,5}$
V. Velovi\' c,$^{6}$
P. Kavanagh,$^{7}$
N. I. Maxted,$^{1,8}$
\newauthor
D. Uro\v sevi\' c,$^{6,9}$
G. P. Rowell,$^{10}$
G. F. Wong,$^{1,8}$
B.-Q. For,$^{11,12}$
A. N. O'Brien,$^{1,5}$
\newauthor
T. J. Galvin,$^{1,5,13}$
L. Staveley-Smith,$^{11,12}$
R. P. Norris,$^{1,5}$
T. Jarrett,$^{14}$
R. Kothes,$^{15}$
\newauthor
K. J. Luken,$^{1,5}$
N. Hurley-Walker,$^{13}$
H. Sano,$^{16}$
D. Oni\' c,$^{6}$
S. Dai,$^{5}$
T. G. Pannuti,$^{17}$
\newauthor
N. F. H. Tothill,$^{1}$
E. J. Crawford,$^{1}$
M. Yew,$^{1}$
I. {Boji{\v c}i{\'c}},$^{1}$
H. D\'{e}nes,$^{18}$ 
N. McClure-Griffiths,$^{19}$
\newauthor
S. Gurovich,$^{20}$
and Y. Fukui$^{16}$
\\
$^{1}$Western Sydney University, Locked Bag 1797, Penrith South DC, NSW 1797, Australia\\
$^{2}$Max-Planck-Institut f\"{u}r extraterrestrische Physik, Giessenbachstra\ss e, 85748 Garching, Germany\\
$^{3}$Observatoire Astronomique de Strasbourg, Universit\'e de Strasbourg, CNRS, 11 rue de l'Universit\'e, F-67000 Strasbourg, France\\
$^{4}$Remeis Observatory and ECAP, Universit\"{a}t Erlangen-N\"{u}rnberg, Sternwartstr. 7, 96049 Bamberg, Germany\\
$^{5}$CSIRO Astronomy and Space Sciences, Australia Telescope National Facility, PO Box 76, Epping, NSW 1710, Australia\\
$^{6}$Department of Astronomy, Faculty of Mathematics, University of Belgrade, Studentski trg 16, 11000 Belgrade, Serbia\\
$^{7}$School of Cosmic Physics, Dublin Institute for Advanced Studies, 31 Fitzwillam Place, Dublin 2, Ireland\\
$^{8}$School of Science, University of New South Wales, Australian Defence Force Academy, Canberra, ACT 2600, Australia\\
$^{9}$Isaac Newton Institute of Chile, Yugoslavia Branch\\
$^{10}$School of Physical Sciences, The University of Adelaide, Adelaide 5005, Australia\\
$^{11}$International Centre for Radio Astronomy Research, University of Western Australia, 35 Stirling Hwy, Crawley, WA 6009, Australia\\ 
$^{12}$ARC Centre of Excellence for All Sky Astrophysics in 3 Dimensions (ASTRO 3D) \\ 
$^{13}$International Centre for Radio Astronomy Research, Curtin University, Bentley, WA 6102, Australia\\
$^{14}$Astrophysics, Cosmology and Gravity Centre (ACGC), Astronomy Department, University of Cape Town, Private Bag X3,\\ Rondebosch 7701, South Africa\\
$^{15}$Dominion Radio Astrophysical Observatory, Herzberg Programs in Astronomy and Astrophysics, National Research Council Canada,\\ PO Box 248, Penticton, BC V2A 6J9, Canada \\
$^{16}$Institute for Advanced Research, Nagoya University, Chikusa-ku, Nagoya 464-8601, Japan\\
$^{17}$Space Science Center, Department of Earth and Space Sciences, Morehead State University, Morehead, KY 40351, USA\\
$^{18}$ASTRON - Netherlands Institute for Radio Astronomy, 7991 PD, Dwingeloo, The Netherlands\\
$^{19}$Research School of Astronomy \& Astrophysics, Australian National University, Canberra ACT 2610, Australia\\
$^{20}$Instituto de Astronom\'ia Te\'orica y Experimental - Observatorio Astron\'omico Co\'rdoba (IATE-OAC-UNC-CONICET),\\
Laprida 854, X5000BGR, C\'ordoba, Argentina\\
}

\date{Accepted XXX. Received YYY; in original form ZZZ}

\pubyear{2018}
\begin{document}
\label{firstpage}
\pagerange{\pageref{firstpage}--\pageref{lastpage}}
\maketitle

\begin{abstract}

We report the discovery of a new Small Magellanic Cloud Pulsar Wind Nebula (PWN) at the edge of the Supernova Remnant (SNR) \dem. The pulsar powered object has a cometary morphology  similar to the Galactic PWN analogs PSR\,B1951+32 and `the mouse'. It is travelling supersonically through the interstellar medium. We estimate the Pulsar kick velocity to be in the range of 700--2000\,km\,s$^{-1}$ for an age between 28--10\,kyr. The radio spectral index for this SNR-PWN-pulsar system is flat (--0.29~$\pm$~0.01) consistent with other similar objects. We infer that the putative pulsar has a radio spectral index of --1.8, which is typical for Galactic pulsars. We searched for dispersion measures (DMs) up to 1000\,cm$^{-3}$\,pc but found no convincing candidates with a S/N greater than 8. We produce a polarisation map for this PWN at 5500~MHz and find a mean fractional polarisation of P~$\sim$23\,percent. The X-ray power-law spectrum ($\Gamma\sim$2) is indicative of non-thermal synchrotron emission as is expected from PWN-pulsar system. Finally, we detect \dem\ in  Infrared (IR) bands. Our IR photometric measurements strongly indicate the presence of shocked gas which is expected for SNRs. However, it is unusual to detect such IR emission in a SNR with a supersonic bow-shock PWN. We also find a low-velocity \HI\ cloud of $\sim$107\,km\,s$^{-1}$ which is possibly interacting with \dem. SNR \dem\ is the first confirmed detection of a pulsar-powered bow shock nebula found outside the Galaxy.

\end{abstract}




\section{Introduction}
 
The Small Magellanic Cloud (SMC) is a gas-rich irregular dwarf galaxy orbiting the Milky Way (MW). However, the proper motion of both the SMC and the Large Magellanic Cloud (LMC) is high, with \citet[and references therein]{2007ApJ...668..949B} arguing that it could be on its first passage about the MW. With a current star formation rate (SFR) of 0.021--0.05\,M$_{\odot}$\,yr$^{-1}$ \citep{2018MNRAS.480.2743F}, it is ranked as the second nearest star forming galaxy after the LMC to the MW. The relatively nearby distance of $\sim$60\,kpc \citep{macri2006new} and the low Galactic foreground absorption ($N_{\rm HI}\sim$6~$\times$~10$^{20}$\,cm$^{-2}$) enables the entire source population in the SMC to be studied in X-rays down to a luminosity of $\sim$10$^{33}$\,erg\,s$^{-1}$ \citep{2012A&A...545A.128H}. The recent star formation activity over the last $\sim$50\,Myr has created an environment where supernova remnants (SNRs) and X-ray pulsars are expected to be plentiful. In line with this, a large population of high-mass X-ray binaries (HMXBs) have been discovered and extensively studied by \citet{Haberl2016}. These objects are typically tens of millions of years old and have spin periods ranging from 1 to 1000\,s. However, the younger population of neutron stars (NSs) is still largely missing. 
 
There are only two known young SMC pulsars: an 8.02\,s anomalous X-ray pulsar, CXOU\,J010043.1--721134, with a characteristic age of $\sim$6700\,yr \citep{Tiengo2008}; The second one powers the only confirmed pulsar wind nebula (PWN) in the SMC, inside the large (74.5\,pc diameter) SNR IKT\,16. IKT\,16 hosts a suspected pulsar with a spin-down luminosity ($\dot{E}$)~{$\sim$}10$^{37}$\,erg\,s$^{-1}$ and an age of $\sim$14.7\,kyr \citep{2011A&A...530A.132O,2015A&A...584A..41M}. The PWN within IKT\,16 was discovered with a dedicated \cxo\ observation, which resolved the hard (2--4.5\,keV) X-ray point source seen with \xmm\ into a faint and soft symmetric nebula surrounding a bright and hard point source. The symmetric morphology suggested that the PWN has not yet interacted with the reverse shock of the SNR. IKT\,16 was also detected at radio continuum frequencies with a typical flat spectral energy distribution (SED)\footnote{Defined as $S_{\nu}\propto\nu^{\alpha}$} but without any indication of a point source coinciding with the X-ray point source at the centre of the nebula. 

Other well studied SNRs in the SMC include SXP\,1062 \citep{2012A&A...537L...1H} which hosts a young Be X-ray binary pulsar with a long spin period in the centre of the remnant. \citet{2014AJ....148...99C} investigated the SNR HFPK\,334 which has a prominent X-ray point source close to the centre of the SNR. This X-ray source was considered to be a background object and not associated with the SNR. At the same time \citet{2015ApJ...803..106R} noted a lack of confirmed and prominent type\,Ia SNRs, indicating that the sample of $\sim$25 SNRs in the SMC may not be representative for any normal galaxy. However and recently, Maggi~et~al. (in prep) list three likely type\,Ia SNRs in the SMC. 


\citet{2017ApJS..230....2B} showed that only five PWNe (out of some 60 confirmed SNRs) are detected so far in the LMC. A further instance was investigated for a relationship between the LMC SNR\,J0529--6653 and the pulsar B0529--66 \citep{2012MNRAS.420.2588B,2017ApJS..230....2B} where the displacement between the midpoint (geometric centre) of the SNR to a pulsar candidate would be consistent with typical kick velocities. However, there was no radio frequency trail of the kind seen in other SNR/PWN systems.

Pulsars and their PWNe, moving supersonically through the ambient medium, are characterised by bow shaped shocks around the pulsar and/or cometary tails \citep{2017SSRv..207..175R}. This population is characterised by Rotation-Powered Pulsars (RPPs) with ages between 10\,kyr--3\,Myr, with a spin-down power ($\dot{E}$) ranging from 10$^{33}$--10$^{37}$\,erg\,s$^{-1}$ \citep{Kargaltsev2017}. Currently, there are $\sim$28 known pulsars which show indications of supersonic motion (most with measured velocities). All of them reside in our Galaxy and exhibit a wide range of morphology \citep[c.f.:][]{2014IJMPS..2860172P,2014A&A...562A.122P,Kargaltsev2017,2019MNRAS.484.4760B}. Although the LMC pulsar J0537--6910 (a.k.a SNR\,N157\,B or 30\,Dor\,B) shows indications of some supersonic motion, its nature is yet to be firmly established. Similarly, LMC SNR\,N\,206 \citep{2002AJ....124.2135K} shows an interesting linear tail-like feature but without an identifiable point source (pulsar) association. The other well established PWNe in the LMC do not exhibit a bow shock morphology, but a typically flat radio SED is evident and located more or less centrally within the remnant \citep{2012A&A...543A.154H}. 

\begin{figure*}
		\includegraphics[scale=1.4,trim=0 0 0 0, clip]{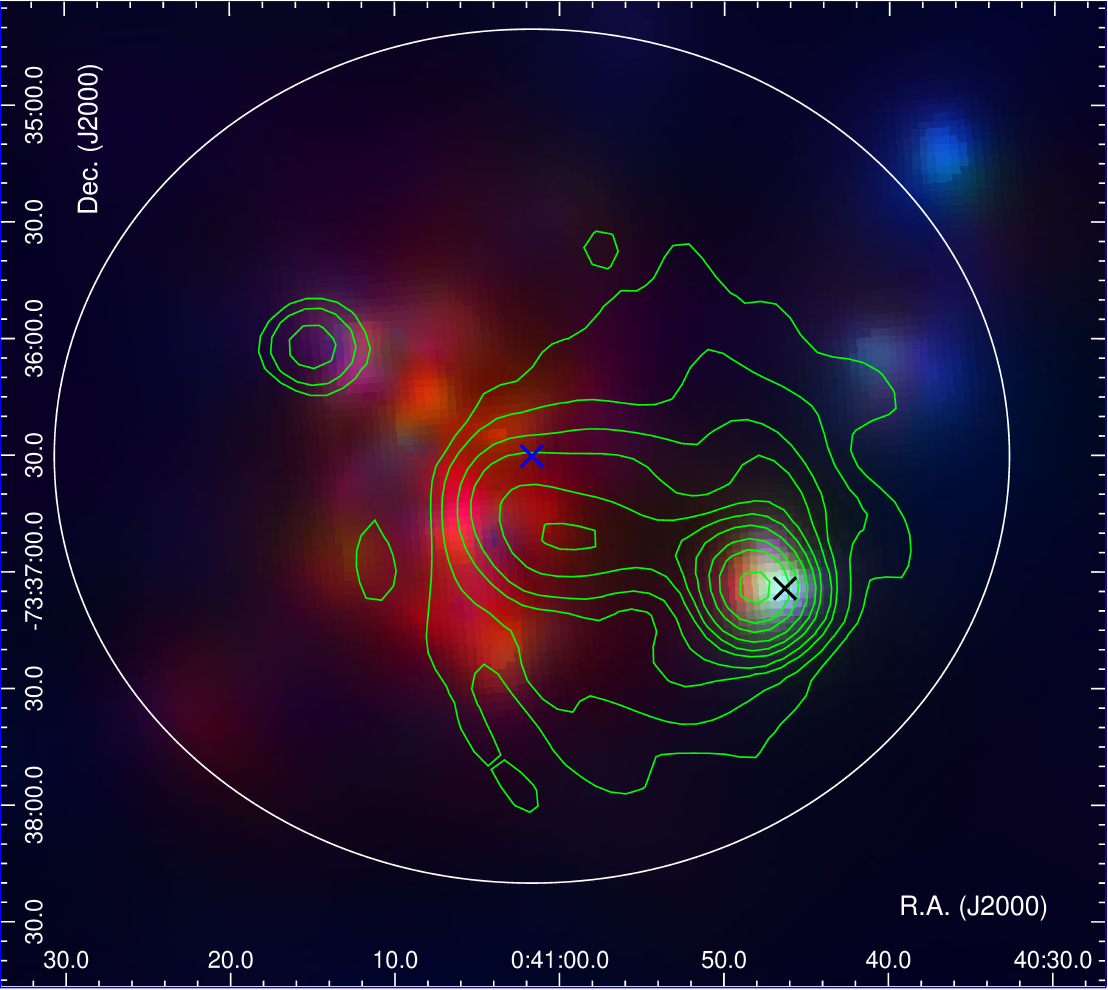}
	\caption{\xmm\ EPIC RGB (R=~0.2--1\,keV, G=~1--2\,keV, B=~2--4.5\,keV) image of \dem. The optical size of the SNR and the centre are marked with a white ellipse (245.5~$\times~$219.7\,arcsec extent) and blue cross respectively \citep{2008A&A...485...63F}. The green contours denote the 1320~MHz radio image. The contours are: 0.5, 1, 2, 3, 5, 7, 10, 15, 20, and 30\,mJy\,beam$^{-1}$ where we measure the local RMS (or 1$\sigma$) to be $\sim$0.1\,mJy\,beam$^{-1}$. The beam size of the radio image is 16.3~$\times$~15.1\,arcsec. The position of the hard X-ray point source (2--4.5\,keV) is marked with a black cross.} 
	\label{fig1}
\end{figure*}

\dem\ (MCSNR\,J0041--7336; HFPK530) is among the largest SNRs in the SMC with a size of 245.5~$\times$~219.7\,arcsec (71.4~$\times$~63.9\,pc at the distance of 60\,kpc to the SMC; see Fig.~\ref{fig1}), is X-ray faint, and has a complex shell like structure. The size of \dem\ is measured from optical (MCELS; see section~\ref{optical}) images while both X-ray and radio images show a smaller extent. With the present sensitivity (and resolution) of our X-ray and radio observations we cannot see the complete shell of SNR \dem, specifically, we are unable to measure any emission in the north and north-west region of the SNR. Therefore the size of the complete shell can only be reliably measured from the optical MCELS images. While lack of sensitivity may play an important role, this could also indicate that the outer regions seen in \OIII\ as radiative shocks contains gas too cool to emit X-rays or radio continuum. A previous \xmm\ study of \dem\ based on observations in which the source was substantially off-axis ($\sim$8\,arcmin) yielded poor constraints on the properties of the SNR due to statistical limitations \citep{2008A&A...485...63F}. However, an interesting outcome of the study was the detection of a soft X-ray source $\sim$67.5\,arcsec from the centre at RA(J2000)=~00$^{h}$40$^{m}$47.7$^{s}$, and DEC(J2000)=~--73$^\circ$37$\arcmin$03$\arcsec$. \citet{2008A&A...485...63F} noted then that the hard X-ray emission also coincided with the peak of the radio continuum image obtained with Australia Telescope Compact Array (ATCA).  

\begin{figure}
	\includegraphics[angle=-90,width=\columnwidth,trim=0 0 45 331, clip]{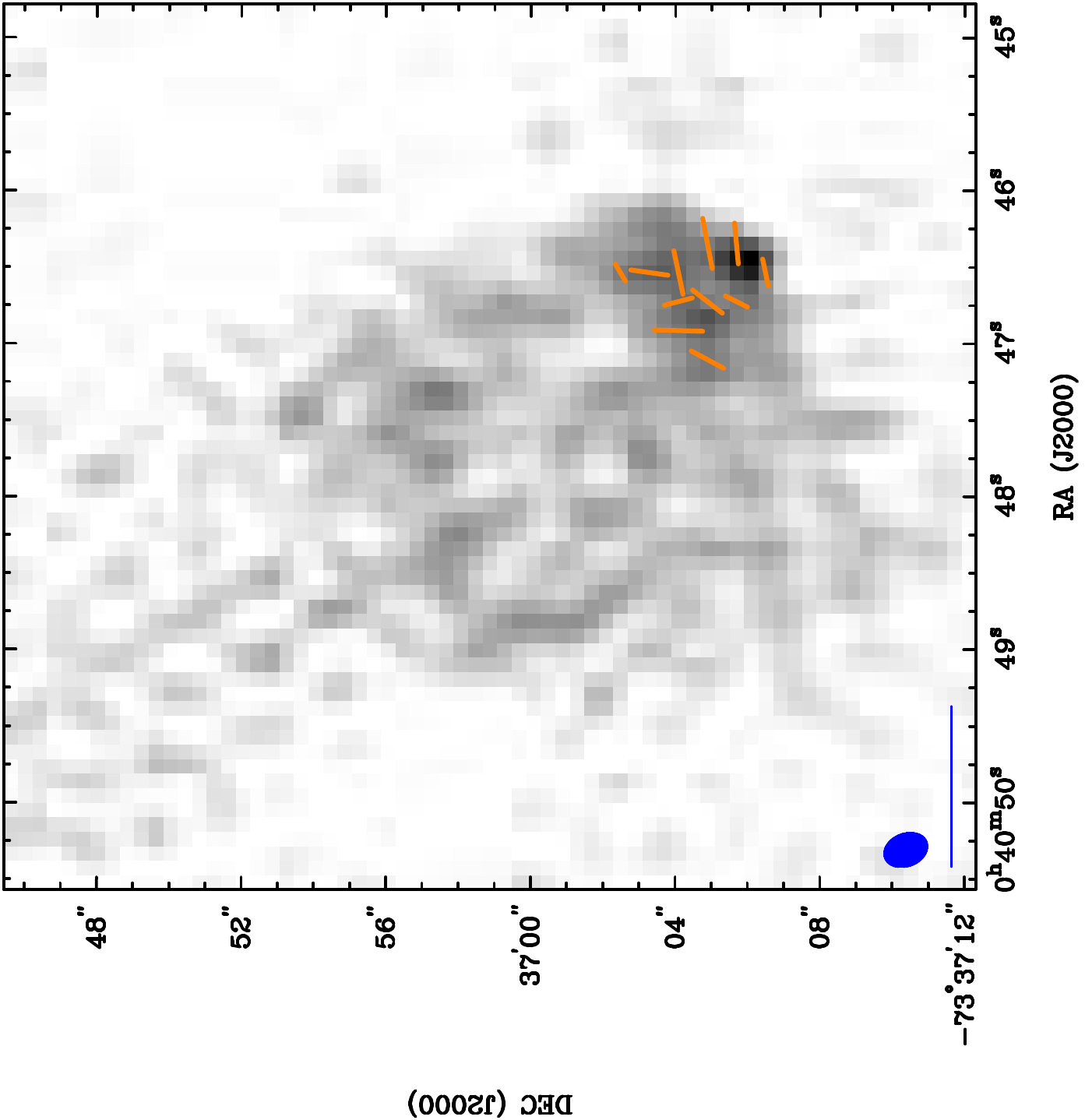}	
	\caption{Fractional polarisation vectors overlaid on the 5500\,MHz ATCA image of \dem\ PWN. The blue ellipse in the lower left corner represents the synthesised beamwidth of 1.2~$\times$~0.89\,arcsec and the blue line below the ellipse represents a polarisation vector of 100\,percent. We used \mbox{{\texttt robust=~--1}} weighting scheme to make this image. The peak fractional polarisation value is P=~32~$\pm$~7\,percent while the average polarisation is measured to be $\sim$23\,percent.}
	\label{fig2}
\end{figure}

\begin{figure*}
	\includegraphics[scale=0.9,angle=-90, trim=33 95 0 0, clip]{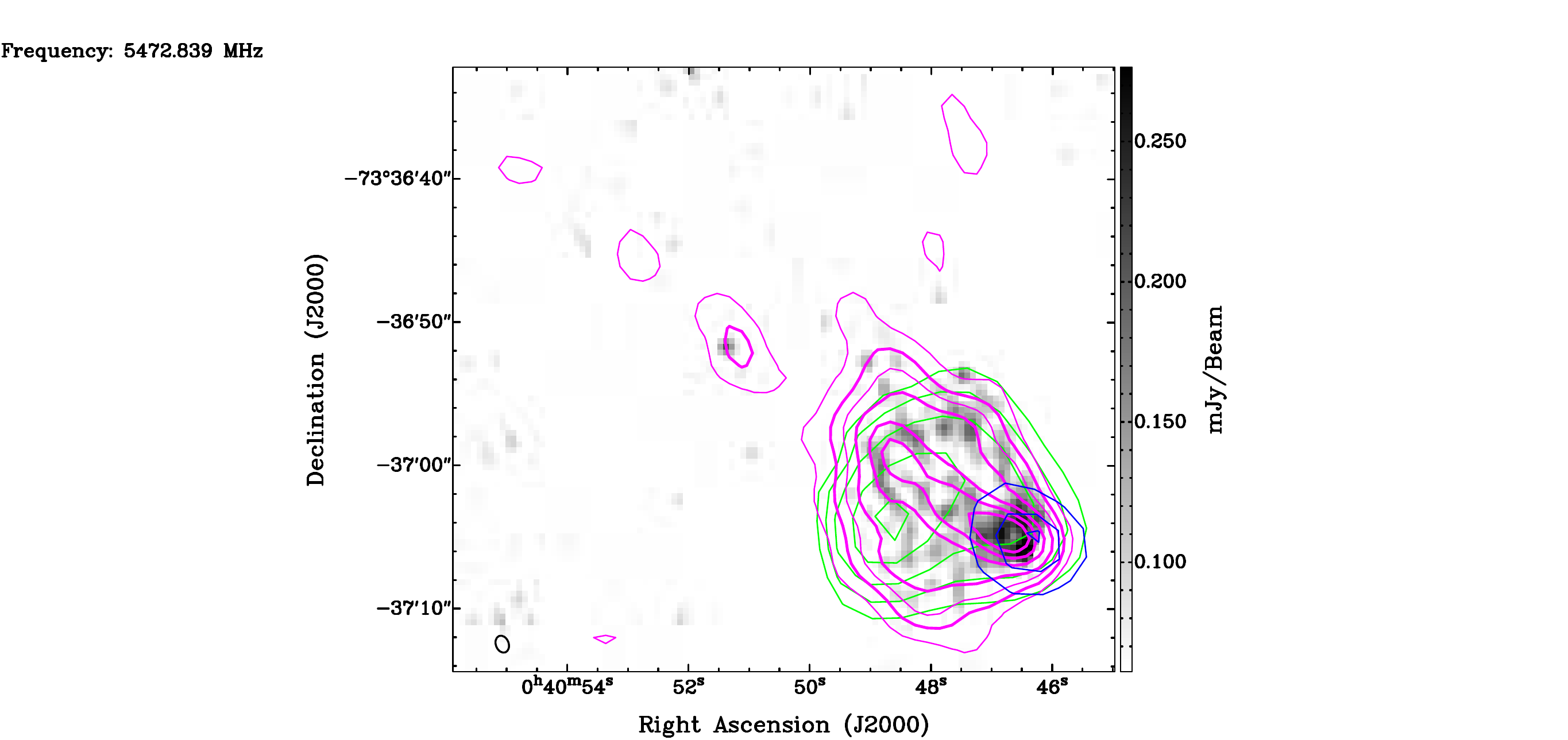}
	\caption{A zoomed in image of the point-like source (pulsar) and PWN within \dem\ showing a comparison between the radio and X-ray wavelengths. The gray-scale image is from the ATCA observation at 5500\,MHz, magenta contours are from the 2100\,MHz image (0.3, 0.5, 0.7, 1, 1.5, 2, and 2.5\,mJy\,beam$^{-1}$); green contours are from \xmm\ (0.2--1\,keV; 1.7, 1.9, 2.1, 2.3, and 2.5 $\times$~10$^{-5}$\,cts\,s$^{-1}$\,pixel$^{-1}$); and blue contours are the \xmm\ (2--4.5\,keV; levels are 1.1, 1.3, and 1.4 $\times$~10$^{-5}$\,cts\,s$^{-1}$\,pixel$^{-1}$). The black ellipse in the lower left corner represents the synthesised beam width of 1.2~$\times$~0.89\,arcsec. An RMS noise at 5500\,MHz image is 0.013\,mJy\,beam$^{-1}$ and 0.1\,mJy\,beam$^{-1}$ at 2100\,MHz image.}
	\label{cabb}
\end{figure*}

\begin{figure}
\centering
\includegraphics[scale=0.35,trim=35 0 0 0, clip]{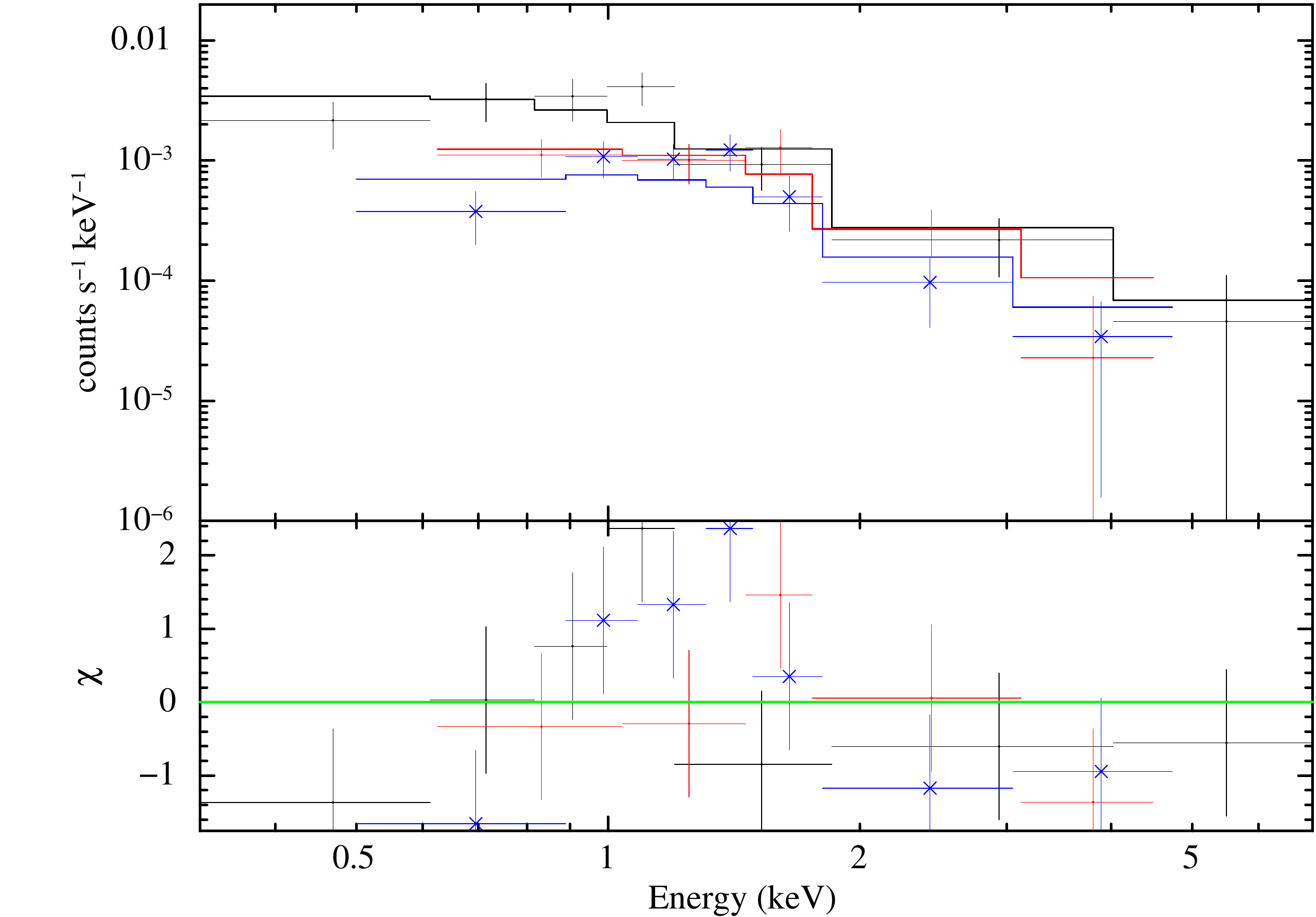}
\caption{The upper panel shows the simultaneous spectral fit using spectra from all the available \xmm/EPIC cameras (PN: ObsID 0301170301 in black, MOS2: ObsID 0301170301 in red and MOS2: ObsID 0764780201 in blue double crosses) along with the best-fit model. The lower panel displays the residuals after the fit. The plots have been rebinned for visual clarity.}
\label{figspec}
\end{figure}

\begin{figure}
\centering
\includegraphics[width=0.99\columnwidth, trim=70 0 35 0, clip]{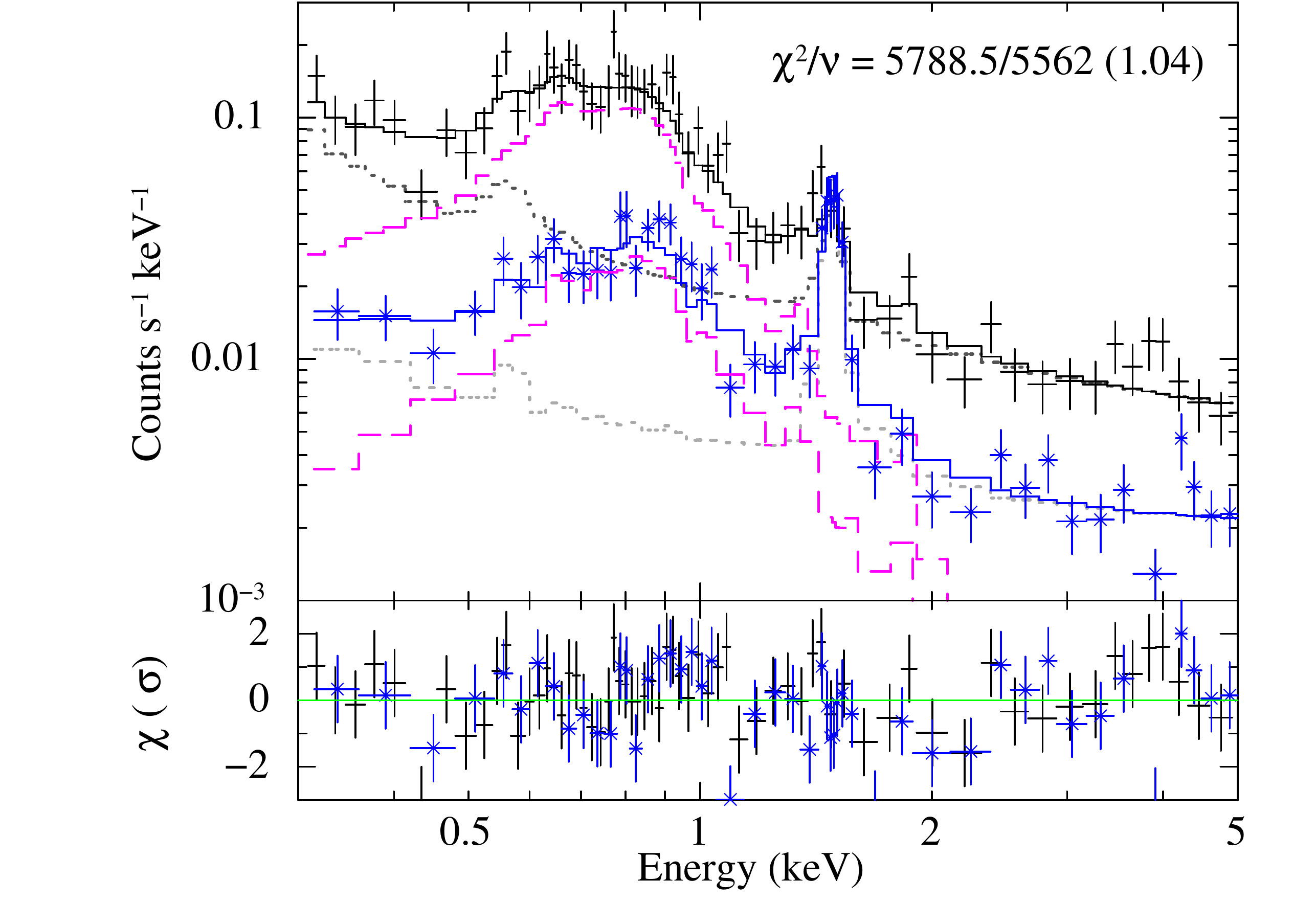}
\caption{X-ray spectra of the diffuse emission of SNR DEM S5. Colours for data are the same as in Fig.\,\ref{figspec}, but MOS data for ObsID 0301170301 have been omitted for the sake of clarity. The total background models for PN and MOS are shown by the dotted gray lines. The SNR emission models  are shown by the dashed magenta lines, as convolved with the PN and MOS spectral responses (top and bottom curves, respectively). The lower panel displays the residuals.}
\label{figspec_diffuse}
\end{figure}

\begin{figure*}
	\includegraphics[width=\textwidth,trim=50 65 75 40,clip]{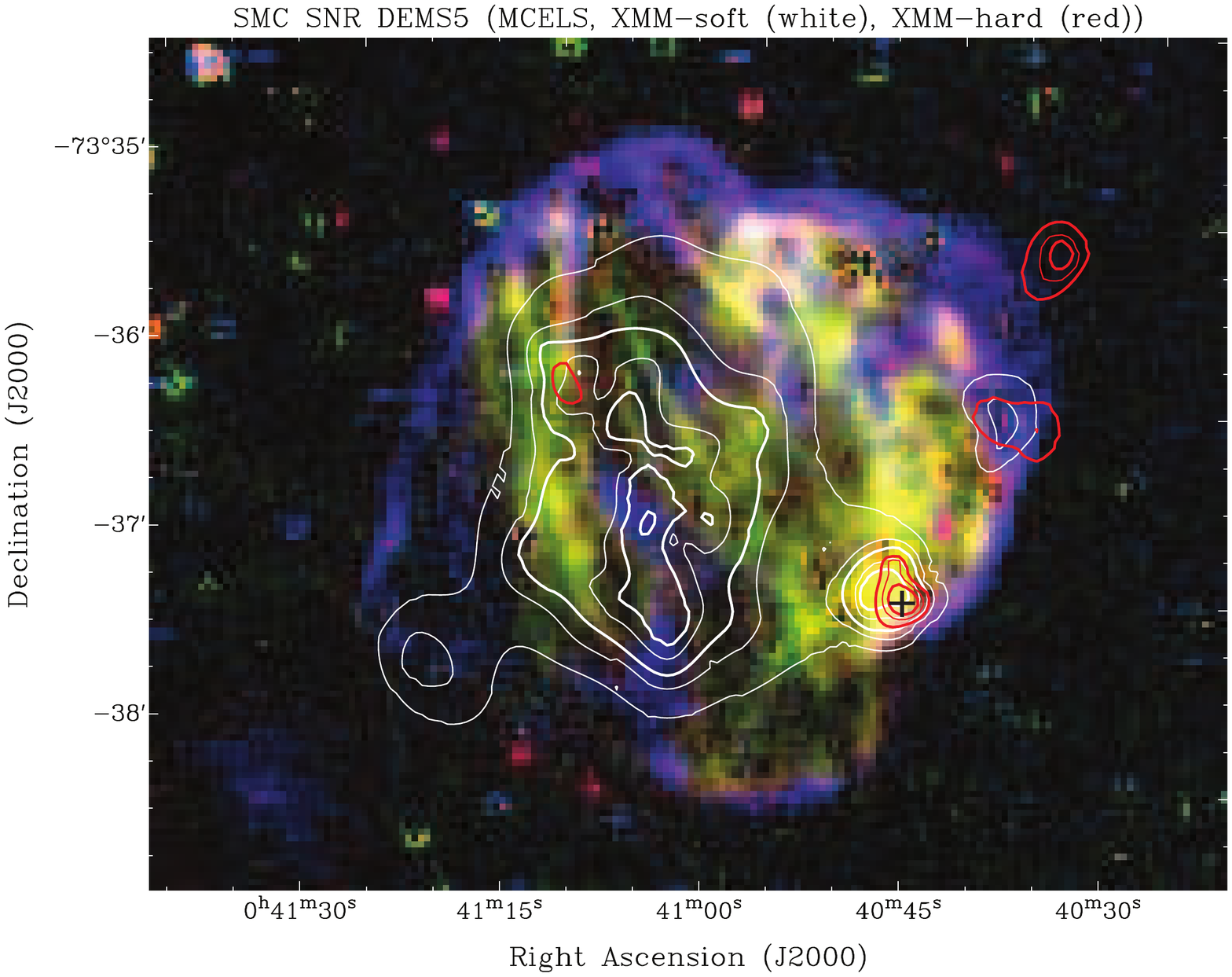}
	\caption{MCELS composite image of \dem\ (RGB=~\Halpha, \SII, \OIII) overlaid with \xmm\ EPIC contours, soft (0.2--1\,keV) in white (levels are 0.5, 0.75, 1, 1.5, and 2~$\times$~10$^{-5}$\,cts\,s$^{-1}$\,pixel$^{-1}$) and hard (2--4.5\,keV) in red (levels are 0.5, 0.75, and 1~$\times$~10$^{-5}$\,cts\,s$^{-1}$\,pixel$^{-1}$). The position of the suggested pulsar is marked with a black plus sign.}
	\label{fig5}
\end{figure*}

Confirmation of the nature of this source is essential to refine our knowledge of pulsars in the SMC and more generally for stellar population studies of dwarf irregular galaxies. In particular, the detection of a new bow shock PWN would be of critical importance to study this small subsample of PWN systems. Furthermore, since the star formation history and metallicity of the SMC is different from that of our Galaxy, it is an interesting system for studying the pulsar population, and comparing the younger population with the plentiful older ones (HMXBs).

In this paper we present an analysis of radio and X-ray observational data of \dem\, as follows: Section~\ref{Sect:data} details the radio X-ray and supplementary optical \& Infrared (IR) data; Sections~\ref{Sect:results} and~\ref{Sect:discussion} respectively contain our results and a discussion. Finally, concluding remarks are given in Section~\ref{Sect:conclusion}.

\section{Observations and Data}
\label{Sect:data}

\subsection{Radio continuum observations}  
We used ATCA observational data from projects CX310 and CX403 that were taken on the 2014 December~30$^\mathrm{th}$ and 31$^\mathrm{st}$; 2015~January~4$^\mathrm{th}$ and 2017~December~22$^\mathrm{nd}$. These observations used the Compact Array Broadband Backend (CABB) (with 2048 MHz bandwidth) in the 6A and 6C array configuration, centred at wavelengths of 3/6\,cm  ($\nu$=~4500--6500 and 8000--10000\,MHz; centred at 5500 and 9000\,MHz), and 13\,cm ($\nu$=~2100\,MHz). The observations were carried out in a frequency switching mode, totaling $\sim$444\,minutes of integration at 4500--6500 and 8000--10000\,MHz bands and 153\,minutes of integration at 2100\,MHz frequency. Source PKS\,B1934--638 was used as the primary (flux density) calibrator for the 2014~December~30$^\mathrm{th}$, 2014~December~31$^\mathrm{th}$, and 2015~January~4$^\mathrm{th}$ data; and source PKS\,B0252--712 for the 2017~December 22$^\mathrm{nd}$ data. Source PKS\,B0230--790 was used as the secondary (phase) calibrator for all four observations (Table~\ref{tab1}). The \textsc{miriad}\footnote{http://www.atnf.csiro.au/computing/software/miriad/} \citep{1995ASPC...77..433S} and \textsc{karma}\footnote{http://www.atnf.csiro.au/computing/software/karma/} \citep{1995ASPC...77..144G} software packages were used to reduce and analyze the data. Images were formed using \textsc{miriad} and the multi-frequency synthesis tasks \citep{1994A&AS..108..585S} therein with a Briggs weighting robust=~--1 parameter for the 5500\,MHz image and 0 for the 2100\,MHz image. Both images were deconvolved by applying a primary beam correction. The same procedure was used for both the \textit{Q} and \textit{U} stokes parameter maps (see Fig.~\ref{fig2}). 

The 2100\,MHz image has a resolution of 3.7~$\times$~2.9\,arcsec with position angle (PA)=~58.1\,deg, while the 5500\,MHz image had a resolution of 1.2~$\times$~0.89\,arcsec and PA=~21.3\,deg. Because of the low surface brightness and lack of short spacings, we were unable to create a reliable image at 9000\,MHz to sample the entire \dem\ extended emission. However, we use the 9000\,MHz image for studying point sources within the field as they are largely not affected by the above shortcomings. Our new images at 2100, 5500, and 9000\,MHz have an RMS noise of 0.1, 0.013, and 0.012\,mJy\,beam$^{-1}$ respectively.

We measured the integrated flux density of \dem\ PWN from 11 separate images spanning a frequency range of 88 to 8640\,MHz, which are summarised in Table~\ref{tbl-flux}. At the same time, we were able to measure the flux density from our two new CABB observations (5500 and 9000\,MHz) of the two point-like sources in the field of \dem\ which we associate with the putative pulsar and an unrelated background source (RA(J2000)=~00$^{h}$41$^{m}$14.82$^{s}$, and DEC(J2000)=~--73\degr35\arcmin59.9\arcsec). 

Similar to the studies of the radio continuum from the LMC by \citet{2017ApJS..230....2B} and Maggi~et~al.~(2018; in prep), we have used all available radio continuum data described in previous SMC surveys \citep{1997A&AS..121..321F,1998A&AS..130..421F,1998PASA...15..280T,2002MNRAS.335.1085F,2004MNRAS.355...44P,2005MNRAS.364..217F,2006MNRAS.367.1379R,2007MNRAS.376.1793P,2011SerAJ.182...43W,2011SerAJ.183..103W,2011SerAJ.183...95C,2012SerAJ.184...93W,2012SerAJ.185...53W}. These data have been taken with the ATCA, Parkes and the Molonglo Synthesis Telescope (MOST). Additionally, we make use of Murchison Widefield Array (MWA) data from \citet{2018MNRAS.480.2743F} and Joseph~et~al.~(2019, in prep.) obtained from the Australia Square Kilometre Array Pathfinder (ASKAP) Early Science Project observations of the SMC.

In Fig.~\ref{cabb} we show the new ATCA radio continuum images at 5500 and 2100\,MHz with the X-ray soft (0.2--1.0\,keV) and hard (2--4.5\,keV) band contours overlaid, obtained from \xmm. For all radio surveys (except the MWA), we used the \textsc{miriad} task \texttt{imfit} in order to extract source integrated flux density, extensions (diameter/axes (D) and PA). Errors in flux density measurements predominately arose from uncertainties in defining the `edge' of the remnant. However, we estimate these errors to be $<$~10\,percent. 

To estimate the flux density from the MWA surveys, we cut out $\approx$3~$\times$~3\,deg areas around the source from the GLEAM wide-band extragalactic catalogue images \citep{2017MNRAS.464.1146H,2018MNRAS.480.2743F}\footnote{http://gleam-vo.icrar.org}. Using the Aegean Tools package \citep{2012MNRAS.422.1812H,2018PASA...35...11H}\footnote{https://github.com/PaulHancock/Aegean}, we ran the Background And Noise Estimator and the Aegean source-finder on the 170--231\,MHz image to detect the SNR/PWNe. We then used the measured position as a prior and allowed the flux to vary when finding it in the 72--103, 103--134, and 139--170\,MHz images (`priorized fitting'). The errors on the integrated flux densities take into account the local RMS, the fitting errors, and the overall flux scale calibration \citep[13 percent at this low Declination]{2017MNRAS.464.1146H}. The source \dem\ was unresolved at all MWA frequencies (as expected) so the integrated flux density is identical to the (background-subtracted) peak flux density.

\subsection{X-ray observations and analysis}
\label{xray}

\dem\ was observed serendipitously with \xmm\ twice between 2006 and 2015 albeit at substantial off-axis angles. The observation details are given in Table~\ref{tabxray}. Fig.~\ref{fig1} displays the \xmm\ EPIC RGB image of \dem, showing the supernova remnant in X-rays and a hard point-like X-ray source shown with a black cross in the south-west direction at a distance of 21.2\,pc from the geometrical centre of the remnant. During both observations the point-like X-ray source was located on a malfunctioning CCD of MOS1 and in the case of ObsID 0764780201, the source was also outside the readout area of the PN operated in small window mode. \xmm/EPIC \citep{2001A&A...365L..18S,2001A&A...365L..27T} observations were processed with the \xmm, data analysis software SAS version 16.1.0\footnote{Science Analysis Software (SAS): http://xmm.esac.esa.int/sas/}.

We searched for periods of high background flaring activity by extracting light curves in the energy range of 7.0--15.0\,keV and removed the time intervals with background rates $\geq$~8 and 2.5\,cts\,ks$^{-1}$~arcmin$^{-2}$ for EPIC-PN and EPIC-MOS respectively \citep{2013A&A...558A...3S}. Events were extracted using the SAS task \texttt{evselect} by applying the standard filtering criteria (\texttt{\#XMMEA\_EP \&\& PATTERN<=4} for EPIC-PN and \texttt{\#XMMEA\_EM \&\& PATTERN<=12} for EPIC-MOS).

In order to investigate whether the hard X-ray point-like source is extended, we extracted a radial profile centered on the source in the energy range of 1--2\,keV from the EPIC-PN image (ObsID 0301170301) as the signal to noise ratio peaks in this energy range. This was fitted with a Gaussian profile and the derived width was compared to that of the \xmm\, Point Spread Function (PSF) at the source position. The radial profile of the hard X-ray source is consistent with the PSF and we derived a 3$\sigma$ upper limit of 14\,arcsec for the X-ray extent of the PWN. To further investigate the presence of a diffuse soft X-ray emission that may be associated with the source (and as may be indicated from Fig.~\ref{cabb}, green contours), we performed the same exercise of comparison with the PSF in the energy range of 0.5--1\,keV. An additional soft X-ray emission is detected peaking at $\sim$78\,arcsec, which originates from the X-ray SNR with its peak in emission at this distance (see X-ray image in Fig.~\ref{fig1}).

We extracted the spectrum of the hard X-ray point source using a circular region of radius 20\,arcsec centered around the source position obtained by \cite{2008A&A...485...63F}. 
The background was chosen from two circular regions of radius 20\,arcsec north and south of the source extraction region at a similar distance to the centre of the SNR. This should remove the contribution of soft X-ray emission from the SNR \dem, but given the irregular morphology of the SNR some uncertainty remains.

The SAS tasks \texttt{rmfgen} and \texttt{arfgen} were used to create the redistribution matrices and ancillary files for the spectral analysis. The spectrum was binned to achieve a minimum of one count per spectral bin. The spectral analysis was performed using the {\small XSPEC} fitting package, version~12.9 \citep{1996ASPC..101...17A} using the C-statistic. Errors were estimated at 90\,percent confidence intervals.
The spectra from both observations were fitted simultaneously with absorbed single-component emission models.
Given the low statistics of the spectra power-law (photon index $\Gamma\sim$2), Bremsstrahlung (temperature kT $\sim$1.7\,keV) and APEC plasma emission (kT $\sim$2.1\,keV) yield formally indistinguishable fit quality ($\chi^2$ values of 121.2, 121.1 and 119.9, respectively, for 185\,deg of freedom). The spectra with best-fit absorbed power-law model is shown in Fig.~\ref{figspec}. 
The index of the emitting particles (\textit{p}) has been calculated by using the equation~\textit{p}=~$\Gamma$~$\times$~2~--~1, where $\Gamma$ is the index of the (synchrotron) energy spectrum. the \textit{p} value is consistent with the expected range of 1~$\leq$~\textit{p}~$\leq$~3 for non-thermal emission. The spectral parameters are tabulated in Table~\ref{tabspec}. No excess absorption was required apart from the Galactic value along the line of sight to the SMC. This is in further support of the source being located at the near side of the SMC and argues against the source being a background object, as the absorption at the source position through the depth of the SMC is N$_{HI}\sim$4~$\times$~10$^{21}$\,cm$^{-2}$ \citep{Stanimiro1999}. The value of $\Gamma$ is consistent with typical X-ray spectral slopes observed in \citep{Kargaltsev2008} as is expected from non-thermal synchrotron emission originating from a PWN/pulsar. We further verified that the X-ray spectral shape and the flux were stable between the two \xmm\ observations separated by 8-years by setting the $\Gamma$ and normalizations free between the observations. A non-varying spectrum is in further support of a PWN (containing a putative pulsar unresolved in the current observations) origin of the source. The detection of a hard X-ray point source with \xmm\, ascertains the fact that we are unable to further resolve the PWN/pulsar composite, and require a higher spatial resolution on-axis {\it Chandra} observation for the purpose.

In parallel, we extracted spectra for the diffuse X-ray emission in the east part of the SNR. For such faint \textit{extended} emission we did not subtract background emission, but rather modeled all sources of background directly in the analysis and added a model for the diffuse emission on top of it. The method we used has been described extensively for \xmm\ in \citet{2016A&A...585A.162M}. Even though we included MOS2 data from a second \xmm\ observation (Table~\ref{tabxray}) not available in the first analysis of \citet{2008A&A...485...63F}, the analysis is still hampered by a relatively short exposure time and off-axis location of the source.

The spectrum of the diffuse emission, shown in Fig.\,\ref{figspec_diffuse}, is obviously thermal, typical for shock-heated plasma in SNRs. We fit the spectrum with a non-equilibrium ionisation shock model with variable abundances (\texttt{vpshock} in XSPEC), finding a temperature $kT = 0.65_{-0.12}^{+0.09}$\,keV and ionisation age $n_e t = 2.2 (_{-0.9}^{+1.3}) \times 10^{11}$\,s\,cm$^{-3}$. 
When let free, the abundances of the main elements (O, Ne, Mg, Fe, and Si) do not depart significantly from the SMC ISM abundance pattern \citep{1992ApJ...384..508R}
With an observed flux in the 0.5--8\,keV band of $(7.6 \pm 0.5) \times 10^{-14}$ \ergcm (absorption-corrected flux of $(1.01_{-0.15}^{+0.40}) \times 10^{-13}$ \ergcm), \dem\ is in the faint end of the SMC SNR population \citep[e.g.][]{2016A&A...585A.162M}.
The best-fit absorption column (on top of the Galactic one) is relatively low ($N_H = 1.1 \times 10^{21}$\,cm$^{-2}$) and at the $3\sigma$ level consistent with 0, as for the PWN candidate. Therefore, from the point of view of absorption, both the hard X-ray source and the diffuse emission are consistent within one another.

Naively, the shape and location of this emission would suggest that it is thermal emission from the supernova ejecta which was shocked and heated by the reverse shock. However, this is not sustained by the data, which are fully consistent with SMC abundances, suggesting instead a shocked ISM origin. It remains possible that shock-heated ejecta contribute to the diffuse emission, at a level low enough not to be significantly detected in the available \xmm\ data.


\begin{table*}
\caption{Summary of ATCA observations for \dem.}
\begin{tabular}{@{}lcccclclc@{}}
\hline
Date       &  Project  &  Array      &Integrated time&   Flux density  & Phase      &  Channels & Frequency   & Bandwidth\\
           &    code   &configuration&  (minutes)    &   calibration   &                                               calibration&           &  (MHz)      &  (MHz)   \\
\hline          
2014-12-30&   CX310    &   6A        &      29.3     &   1934--638    & 0230--790 &     2049   & 5500, 9000  &  2048    \\ 
2014-12-31&   CX310    &   6A        &      153      &   1934--638    & 0230--790 &     2049   &    2100     &  2048    \\
2015-01-04&   CX310    &   6A        &      37.3     &   1934--638    & 0230--790 &    2049    & 5500, 9000  &  2048    \\
2017-12-22&   CX403    &   6C        &     377.3     &   0252--712    & 0230--790 &    2049    & 5500, 9000  & 2048     \\
\hline
\label{tab1}
\end{tabular}
\end{table*}


\begin{table*}
\caption{Measurements of integrated flux density of \dem\ PWN and putative pulsar as well as the unrelated background source found within the field of \dem. Because of the poor survey resolution and sensitivity we couldn't detect or measure the flux density of the pulsar and background source at MWA frequencies.}
\begin{tabular}{@{}lcclll@{}}
\hline
$\nu$ & PWN $S_{\nu}$   & Pulsar $S_{\nu}$    & Background $S_{\nu}$ & Telescope & Reference\\
(MHz) &  (Jy)           & (Jy)                & (Jy)                 &           &          \\
\hline
88    & $0.288\pm0.089$ &                     &                      & MWA   & \citet{2018MNRAS.480.2743F}\\
118   & $0.256\pm0.052$ &                     &                      & MWA   & \citet{2018MNRAS.480.2743F}\\
155   & $0.234\pm0.040$ &                     &                      & MWA   & \citet{2018MNRAS.480.2743F}\\
200   & $0.205\pm0.033$ &                     &                      & MWA   & \citet{2018MNRAS.480.2743F}\\
843   & $0.150\pm0.015$ &                     & $0.0044\pm0.0005$    & MOST  & Maggi et. al.  (in prep.)\\
960   & $0.138\pm0.014$ &                     & $0.0038\pm0.0004$    & ASKAP & Joseph et. al. (in prep.)\\
1320  & $0.130\pm0.013$ &                     & $0.0028\pm0.0003$    & ASKAP & Joseph et. al. (in prep.)\\
1420  & $0.128\pm0.013$ &                     &                      & ATCA  & Maggi et. al. (in prep.)\\
2100  & --            & $0.003\pm0.001$     & $0.0017\pm0.0002$    & ATCA  & This work \\
2370  & $0.114\pm0.011$ &                     &                      & ATCA  & Maggi et. al. (in prep.)\\
4800  & $0.085\pm0.009$ &                     &                      & ATCA  & Maggi et. al. (in prep.)\\
5500  & --             & $0.00042\pm0.00003$ & $0.00087\pm0.00003$  & ATCA  & This work \\
8640  & $0.072\pm0.007$ &                     &                      & ATCA  & Maggi et. al. (in prep.)\\
9000  & --             & $0.00022\pm0.00007$ & $0.00065\pm0.00007$  & ATCA  & This work\\
\hline
$\alpha\pm\Delta\alpha$  & $-0.29\pm0.01$ & $-1.8\pm0.2$ & $-0.81\pm0.04$ &   & \\
\hline
\label{tbl-flux}
\end{tabular}
\end{table*}


\begin{table*} 
\caption{ \xmm\ observations of the hard X-ray point source inside \dem.} 
\begin{tabular}{c c c c c c} 
\hline \hline 
Date             & ObsID           & Exposure   & CCD Readout mode & Off-axis angle & Telecope vignetting\\ 
                    &                      & PN / MOS2 &  PN / MOS2              & PN / MOS2      & PN / MOS2 \\
yyyy/mm/dd &                      & (ks)              &                                 &                        &                  \\ 
\hline 
2006/06/04 & 0301170301 & 15.5 / 21.1 & Full Frame / Full Frame       & 8.4\arcmin  / 7.4\arcmin & 0.61 / 0.65 \\ 
2015/10/16 & 0764780201 & --     /  47.9 & Small Window / Full Frame & --  / 9.7\arcmin               & -- / 0.53 \\ 
\hline 
\label{tabxray} 
\end{tabular} 
\end{table*}


\subsection{Optical and Infrared observations}
\label{optical}

The Magellanic Cloud Emission Line Survey (MCELS) was carried out at the 0.6-m University of Michigan/CTIO Curtis Schmidt telescope, equipped with a SITE 2048~$\times$~2048 CCD, which gave a field of 1.35~$\times$~1.35\,deg at a scale of 2.4~$\times$~2.4\,arcsec\,pixel$^{-1}$ \citep{2000ASPC..221...83S,Pellegrini2012}. The SMC was mapped in narrow bands corresponding to \Halpha, \OIII\ ($\lambda$=~5007\,\AA), and \SII\ ($\lambda$=~6716, 6731\,\AA), plus matched red and green continuum bands. All the data have been continuum subtracted, flux-calibrated and assembled into mosaic images. A cutout around the area of \dem\ can be seen in Fig.~\ref{fig5}.

We also make use of SMC observations from {\it Spitzer} SAGE \citep[Surveying the Agents of a Galaxy's Evolution, ][]{2011AJ....142..102G}. Specifically, we used 4 IRAC bands: band-1 (3.6\,$\mu$m), band-2 (4.5\,$\mu$m), band-3 (5.8\,$\mu$m), and band-4 (8.0\,$\mu$m), as well as MIPS at 24\,$\mu$m.

\subsection{\HI\ observations}
\label{s:HIobs}

Observations of~\HI\ have previously been carried out over the whole SMC (20\,deg$^{2}$) by \citet{Stanimiro1999} using the ATCA 375-m configuration and the Parkes telescopes. The angular resolution was 98\,arcsec, and velocity resolution was 
1.65\,km\,s$^{-1}$, with a column density sensitivity of 4.2~$\times$~10$^{18}$\,cm$^{-2}$ in the same velocity interval. The heliocentric reference frame was used to define velocities.

Recent ASKAP data are also available over a similar field-of-view, but at the higher angular resolution of 27~$\times$~35\,arcsec and the lower velocity resolution of 3.9\,km\,s$^{-1}$ \citep{2018NatAs...2..901M}. The column density sensitivity per 3.9\,km\,s$^{-1}$ channel is similar to the ATCA data at 5.2~$\times$~10$^{18}$\,cm$^{-2}$.

\section{Results}
\label{Sect:results}

\subsection{PWN and compact source inside the SNR \dem }
We detect a point-like source at 5500\,MHz, which we consider likely to be a pulsar, at RA(J2000)=~00$^{h}$40$^{m}$46.49$^{s}$, and DEC(J2000)=~--73\degr37\arcmin5.8\arcsec, with a trail/jet behind it and a nebulosity which is morphologically indicative of a PWN (Fig.~\ref{cabb}). This trail is clearly visible only in our 2100 and 1320\,MHz images because of the sensitivity and missing short spacings at higher frequencies. One should also consider that this trail might be a so-called `relic' PWN behind the present putative pulsar similar to Galactic SNR/PWN G5.4--1.2 \citep{Kothes2017}. While somewhat unlikely, as the trail is still part of PWN-system (see Fig.~\ref{fig1}), only future higher sensitivity and resolution radio continuum observations may reveal true nature of this possibility.


The position of the radio point source is also coincident with the hard X-ray point source. There seems to be a small displacement ($<$~2\,arcsec) between the centremost blue (2--4.5\,keV) X-ray contour and the radio point-like source.
This is attributed to the fact that the X-ray counterpart of the putative pulsar cannot be separated from the PWN/pulsar composite with the current \xmm\, observations and hence its accurate X-ray position cannot be ascertained. Fig.~\ref{cabb} further shows that the PWN emission close to the pulsar is dominated by hard X-rays. This is due to the fact that the X-ray synchrotron electrons have a much shorter lifetime than radio ones and the X-ray synchrotron spectrum from the PWN would get steeper when moving further away from the pulsar. 

The measured integrated flux density of the point source (putative pulsar) at 5500 and 9000\,MHz is 0.42\.mJy ($\sim$14$\sigma$) and $<$~0.22\,mJy ($\sim$3$\sigma$) respectively. At 2100\,MHz, we cannot precisely measure flux density, but there is a strong indication of an order of magnitude stronger integrated flux density (around 3\,mJy, see Table~\ref{tbl-flux}). This would make the SED of this point source (putative pulsar) very steep with a spectral index of $\alpha\sim$--1.8. The SED of most known Galactic pulsars can be described above 100\,MHz by a simple power-law with a mean spectral index of $\sim$--1.8 \citep{2016RAA....16..159H}, with a range from --0.46 to --4.84. Therefore, we propose that this point source is most likely the compact remnant (pulsar or a fast spinning NS) of the \dem\ supernova explosion with its associated PWN.  Also, ultra steep spectrum and Compact Steep Spectrum (CSS) sources would have such a steep radio spectrum \citep{2016AN....337...36C}. However, as we see the nebulosity behind the point source, it is very unlikely to be CSS with jet Active Galactic Nuclei
(AGN) like structure. 

%
%

\begin{figure*}
	\includegraphics[scale=0.75,angle=-90,trim=0 30 0 70,clip]{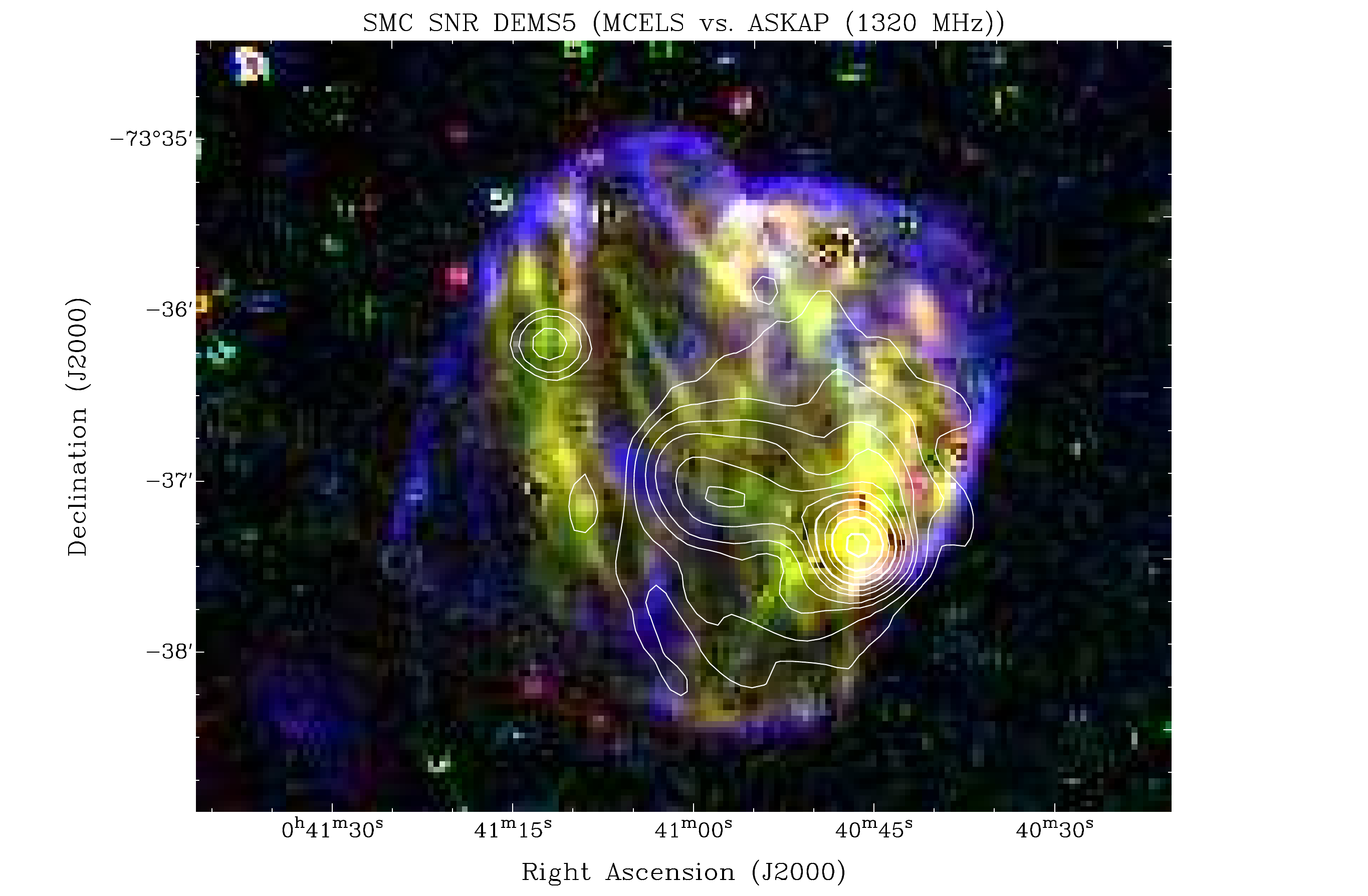}
	\caption{MCELS composite image of \dem\ [HFPK 530] (RGB=~\Halpha, \SII, \OIII) overlaid with ASKAP contours (white solid) at 1320\,MHz. The radio image contours are as same as in Fig.~\ref{fig1}. An obvious radio continuum tail-like feature is clearly pointing towards the geometric centre of \dem. We also note a point source north-east from the remnant centre which we classify as a most likely background galaxy. }
	\label{fig7}
\end{figure*}

Figs.~\ref{fig1}, \ref{cabb} and \ref{fig7} show trail-like emission from PWN with the length of $\sim$37\,arcsec ($\sim$10.8\,pc). The direction of the trail is pointing towards the geometric centre of the SNR (RA(J2000)=~00$^{h}$41$^{m}$01.675$^{s}$ and DEC(J2000)=~--73\degr36\arcmin30.38\arcsec). We also note that the putative pulsar (compact source) is `leading' the conically shaped PWN that trail behind it. This is a strong indication that the putative pulsar is moving away from the SNR explosion site at a supersonic velocity (see Table~\ref{tab4}). Morphologically, this would qualify the \dem\ system (pulsar and PWN) as a `rifle bullet' type \citep{2019MNRAS.484.4760B} where the spin and velocity are aligned. The most likely Galactic bow shock PWN analog would be PSR\,B1951+32 in SNR CTB\,80 \citep[see][]{1995ApJ...439..722S,2004ApJ...610L..33M} and `mouse' PWN (J1747--2956) \citep{2004ApJ...616..383G,2018ApJ...861....5K} where a similar morphology is detected. However, according to \citet{Kargaltsev2017} PSR\,B1951+32 doesn't move supersonically. Fig.~\ref{fig7} clearly shows that the putative pulsar with its PWN is reaching the edge of the optical remnant. The putative pulsar may be even outside of it, if the motion is not entirely perpendicular to the line of sight. Although purely visually, it would be probably just at the edge. The densities outside and inside the SNR should be different, with outside expected to be much higher. Therefore, we expect the PWN to be somewhat narrower with higher ambient density. 

In Table~\ref{tab4}, we used various SNR ages (from 5 to 50\,kyr) to estimate the kick (transverse) velocity of the pulsar. We also varied the distance to \dem\ as \citet{0004-637X-816-2-49} suggested that at the western end of the SMC might reach up to 67.5\,kpc. The observed distance between the radio point-like source (putative pulsar) and the centre of the SNR is measured to be $\sim$73\,arcsec (or 21.2/23.9\,pc at the 60/67.5\,kpc distance respectively and assuming motion to be perpendicular to the line of sight).

The kick (transverse) velocities in other similar PWN bow shock systems are estimated to be in range of 60--2000\,km\,s$^{-1}$ \citep[see their table~1]{Kargaltsev2017} for the pulsar through the ambient medium. We take this velocity as a possible upper limit which then indicates a very high \dem\ pulsar velocity through the ambient medium given that the minimum age would be in order of $>$10\,kyr (see Table~\ref{tab4}). Other constraints on the age of \dem\ come from the IKT\,16 PWN/SNR system which has an estimated age of 14.7\,kyr and has a somewhat larger SNR diameter of 74.5\,pc vs 67.6\,pc for \dem\, indicating that IKT\,16 PWN/SNR is expanding at a somewhat lower rate probably because the local environment is denser than in/around \dem\ ($N_{\rm \HI}\sim$4~$\times$~10$^{21}$\,cm$^{-2}$ for \dem\ vs. $N_{\rm \HI}\sim$6~$\times$~10$^{21}$\,cm$^{-2}$ for IKT\,16). However, one should be  cautious about the line-of-sight column density, calculated by integrating  along several kpc for the SMC because of variations in the local ambient medium density. This would suggest that the age of \dem\ PWN is between 10 and 15\,kyr with kick velocities of $\sim$1500--2500\,km\,s$^{-1}$. Also, \citet{2017SSRv..207..175R} suggest that the pulsar would eventually leave the SNR at an age of $\sim$20--200\,kyr which further limits our \dem\ PWN as it is still within the SNR boundaries. Given a lower age limit of $\sim$10\,kyr, the kick-off velocity of $\sim$2000\,km~s$^{-1}$ is derived using its distance from the \dem\ geometric centre, which is the most likely explosion site in a uniform ISM. If the ISM is inhomogeneous or has a large scale gradient the explosion site would be shifted ($\sim$4.5\,arcsec or 1.3\,pc) from the geometrical centre towards the west (at RA(J2000)=~00$^{h}$41$^{m}$01.53$^{s}$ and DEC(J2000)=~--73\degr36\arcmin26.6\arcsec; with ellipse radius of $\sim$34~$\times~$22\,pc). In any case, a large kick velocity would certainly still need to be accounted for. We emphasize that the age and size of different SNRs might not be appropriate to compare as they all come with their own unique characteristics. 

The fastest known moving pulsars are `Morla' (J0357+3205) \citep{Kargaltsev2017,2013ApJ...765...36M} and `Lighthouse' \citep{2014IJMPS..2860172P,2014A&A...562A.122P,Kargaltsev2017} at $\sim$2000\,km\,s$^{-1}$. We note that our \dem\ PWN cometary tail has a conical shape with the angle behind the putative pulsar of $\sim$75--80\,deg, which suggests somewhat slower velocity than $\sim$2000\,km\,s$^{-1}$ that we estimate above; unless our PWN system is not just characterised as `rifle bullet' but also `frisbee' or `cart-wheel' \citep{2019MNRAS.484.4760B}. Additionally, \citet{Holland2017} and \citet{2018ApJ...856...18K} suggested that pulsars move preferentially opposite to the ejecta, which lends support to the interpretation that the diffuse X-ray emission originates from shock heated ejecta.

\dem\ is a very large SNR in general, which argues for a late stage of evolution. We use \cite{1988ApJ...334..252C} for calculations of radiative SNRs. Using their equations for the onset of the pressure driven snowplow or radiative phase, we obtain a minimum ambient density of 0.35\,cm$^{-3}$ and a lower limit for the age of 24\,kyr, using 22\,pc radius (towards the west) and 10$^{51}$\,erg for the explosion energy -- certainly nothing unusual. For a density of 1\,cm$^{-3}$ the age would be about 48\,kyr but the age estimate could be even a lot higher. We also need to consider the amount of material that would have been swept up in such a large volume if we have normal densities. For this to still be a Sedov SNR we would need very low densities to make this still applicable. The largest distance between the most likely location of the supernova explosion and the edge of the SNR -- as indicated by the \OIII\ shell -- is almost straight to the east. In between, we find the diffuse X-ray emission, which might be including some shock heated ejecta. In this case, the reverse shock must have past this area which stretches all the way to the centre of the SNR. With an assumed ejecta mass of 10 solar masses, a radius of 40\,pc and the canonical value for the explosion energy of 10$^{51}$\,erg, we require an ambient density of 0.02\,cm$^{-3}$ -- if the reverse shock just reached the location of the SN \citep[see their equations]{1995PhR...256..157M}. This is the typical density in the interior of a stellar wind bubble \citep[e.g.][]{1977ApJ...218..377W}. With these parameters the age would be 28\,kyr and the PWN/Pulsar system kick-off velocity would decrease to a `reasonable' 700--800\,km\,s$^{-1}$. We note that the ejecta mass is not very critical here. If we insert this into the equations of \citet{1988ApJ...334..252C} we get an ambient density of 0.45\,cm$^{-3}$ for the right (west) part of the SNR. Therefore, this scenario seems to be at least plausible. This implies that the supernova went off inside a low density bubble close to the edge. Alternatively, if the plasma temperature ($kT=0.8$\,keV represents that of shocked ambient ISM as the limited spectral analysis indicates, a naive Sedov model can be used to set an age $\gtrsim$~15\,kyr, or, taking a non detection in X-ray of the outer, radiative shell ($kT \lesssim0.16$\,keV), an even larger age limit of $\gtrsim$~30\,kyr would be implied.


\subsection {Radio Spectral Index}
\begin{figure}
	\includegraphics[angle=-90,width=\columnwidth]{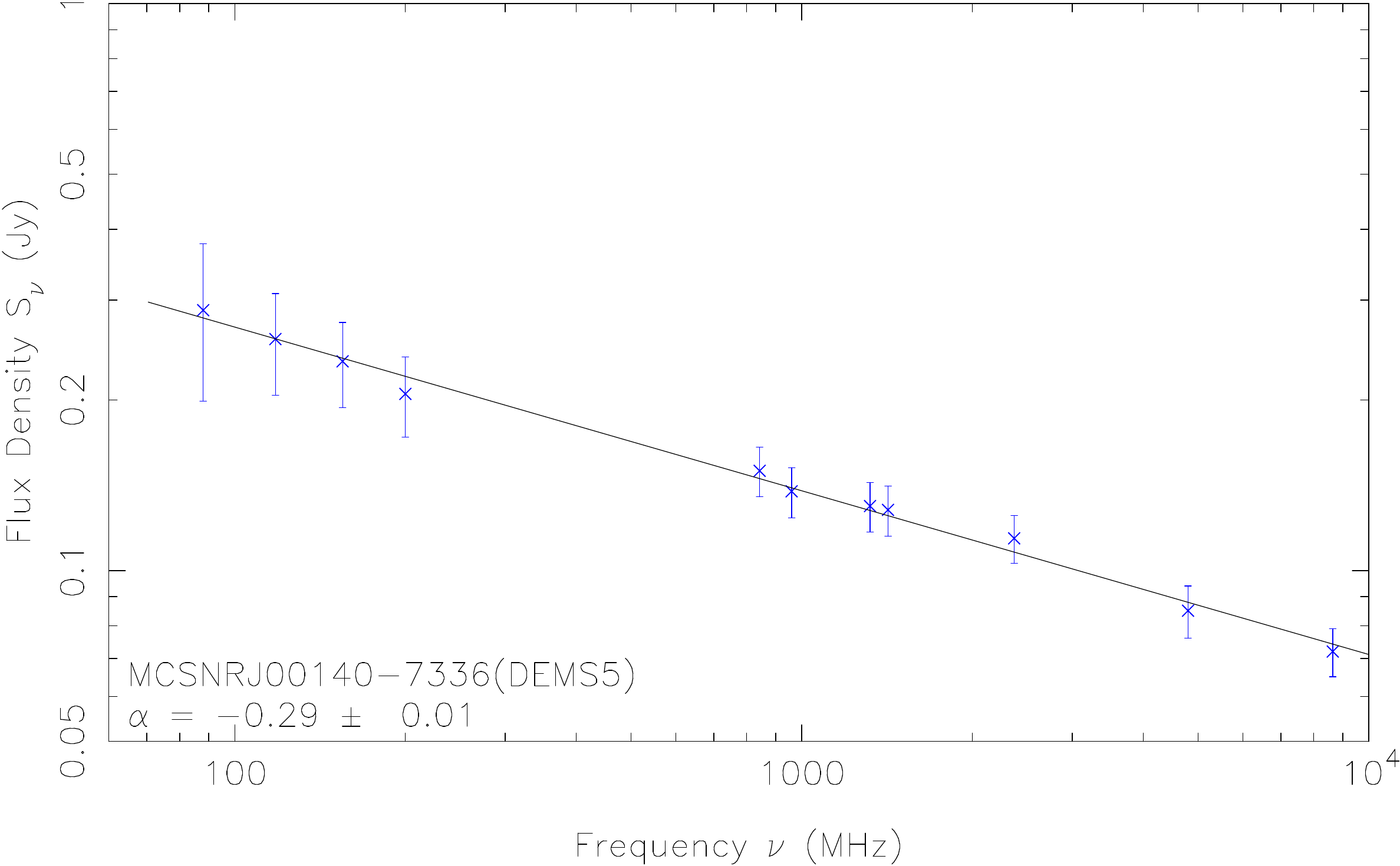}
	\caption{Radio continuum SED of PWN \dem.}
	\label{radioSED}
\end{figure}

Using the flux density measurements from Table~\ref{tbl-flux}, we estimate a spectral index of $\alpha$=~--0.29~$\pm$~0.01 across the remnant (including pulsar and PWN) (Fig.~\ref{radioSED}). The radio spectral index is estimated from the slope of a linear least squares regression, in logarithmic space. From all our radio images it is clear that emission from the putative PWN dominates the radio emission (see section~\ref{polarisation}).

\begin{table}
\caption{X-ray spectral parameters of the point-like source in \dem. Absorption$^b$ was fixed to 6~$\times~$10$^{20}$\,cm$^{-2}$. Errors are quoted at 90\% confidence.}
\begin{tabular}{l c }
\hline \hline
Parameters & Value \\
\hline
 $\Gamma$ & 1.99$^{+0.26}_{-0.28}$ \\
Flux $^a$ (0.5--8.0\,keV) &  3.1~$\pm$~0.9 \\
Flux (unabsorbed) $^a$ (0.3--8.0\,keV) &  3.6~$\pm$~1.0 \\
X-ray Luminosity at $d_{smc}$=~60\,kpc (erg~s$^{-1}$) & 1.5~$\times~$10$^{34}$ \\
X-ray Luminosity at $d_{smc}$=~67.5\,kpc (erg~s$^{-1}$) &1.9~$\times$~10$^{34}$ \\
 \hline
\end{tabular}\\
$^{a}$ Flux in units of 10$^{-14}$ \ergcm\ \\
$^{b}$ Absorption was fixed equal to the Galactic value along the line of sight of SMC to N$_{Hgal}$=~6$\times$10$^{20}$\,cm$^{-2}$ \\
\label{tabspec}
\end{table}

The overall radio continuum spectral index \mbox{$\alpha$=~--0.29~$\pm$~0.01} of the remnant (Fig.~\ref{radioSED}) adds further support to our claim of the PWN nature of the emission. This value of $\alpha$ falls below that of a typical SNR of --0.5 \citep{2017ApJS..230....2B}, and is in line with what is expected from a SNR/PWN system (e.g., the spectral index of the SNR and PWN for LMC SNR\,J0453--6829 is --0.39~$\pm$~0.03; \citet{2012A&A...543A.154H}). As we were unable to disentangle the PWN emission from the rest of the shell, we were not able to estimate the spectral index of the SNR separately. We also note that the estimated value of $\alpha$=~--0.29 could be somewhat flatter as our flux density estimates at lower frequencies includes the emission of the pulsar which might not be negligible but is a minor flux density contributor to the whole \dem\ system (SNR-PWN-Pulsar). We cannot rule out radio emission from the shell but the present radio images show little evidence of any, so we assume that all the radio emission is coming from the putative pulsar (compact source) and its PWN even though we recognise that some emission must come from the SNR itself.

\subsection{Search for the Radio Pulsar at Parkes}
On 2018~September~14{\bf $^\mathrm{th}$} we observed the point-like source for 2700 seconds using the Directors Time of the ATNF Parkes radio telescope in NSW, Australia. The central beam of the multibeam receiver was used and data were recorded using the Parkes Digital Filterbank (PDFB) systems. We used a 2-bit sampling and a sampling time of 64\,microsecond with 512 frequency channels over 256\,MHz bandwidth (centred at 1369\,MHz). The pulsar searching software package PRESTO \citep{2001AAS...19911903R} was used to carry out RFI zapping and masking, de-dispersion and searches of periodic signals. We searched for dispersion measures (DMs) of up to 1000\,cm$^{-3}$\,pc, but found no convincing candidates with a S/N greater than 8. For a pulse period of 10\,ms and DM of 200\,cm$^{-3}$\,pc and assuming a pulse duty cycle of 10\,percent, our observations provide a flux density limit of 0.24\,mJy at 1369\,MHz  \citep[see e.g., equation~9 of][]{2017MNRAS.472.1458D}.

\begin{table}
 \caption{Derived Pulsar kick velocity for the \dem\ progenitor: estimated as a function of SNR age ($V_{\rm psr}$=~2~$\times~d_{\rm smc}~\times$~TAN($\theta/2)$/age) which simplifies to  $(\theta ~\times~d_{\rm smc})$/age since $\theta~<<~1$. The angular distance is $\theta$=~73\,arcsec. The adopted distance of $d_{\rm smc}$=~67.5\,kpc for the western side of the SMC is suggested by \citet{0004-637X-816-2-49}. }
 \begin{tabular}{l c c }
 \hline \hline
 SNR age& $V_{\rm psr\,at\,60\,kpc}$ & $V_{\rm psr\,at\,67.5\,kpc}$\\
 (kyr)  & (km~s$^{-1}$)             & (km~s$^{-1}$)\\ \hline
  5.0   & 4157                      & 4676 \\
  7.5   & 2771                      & 3117 \\
  10.0  & 2078                      & 2338 \\
  12.5  & 1663                      & 1870 \\
  15.0  & 1386                      & 1559 \\
  17.5  & 1188                      & 1336 \\
  20.0  & 1039                      & 1169 \\
  22.5  & 924                       & 1039 \\
  25.0  & 831                       & 935  \\
  30.0  & 693                       & 779  \\
  40.0  & 520                       & 584  \\
  50.0  & 416                       & 468  \\
  \hline
\end{tabular}\\
\label{tab4}
\end{table}

\begin{figure*}
  \subfloat{\includegraphics[width=0.37\textwidth, angle=-90, trim=0 175 0 240, clip]{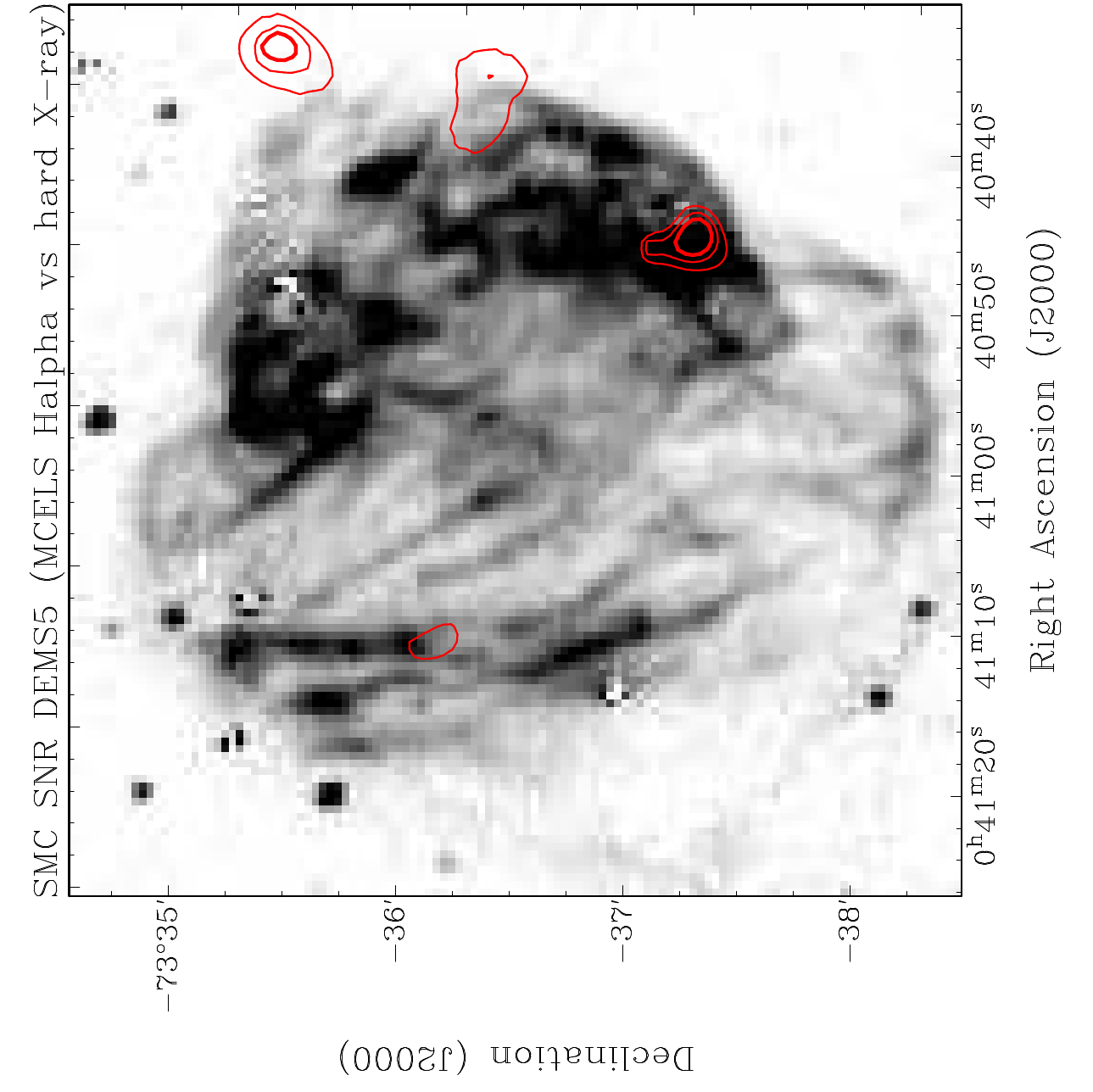}}
  \subfloat{\includegraphics[width=0.37\textwidth, angle=-90, trim=0 240 0 240, clip]{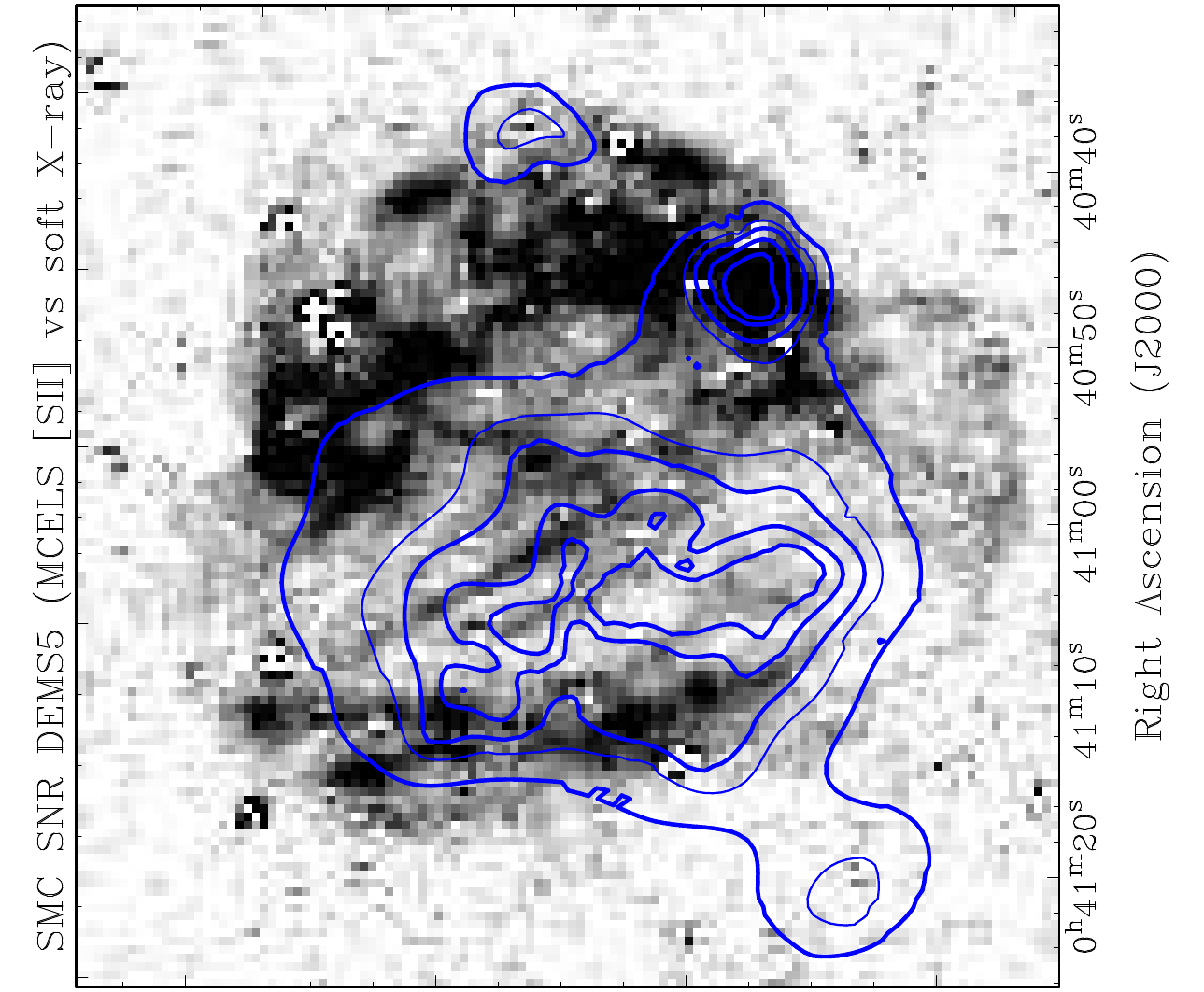}}
  \subfloat{\includegraphics[width=0.37\textwidth, angle=-90, trim=0 240 0 240, clip]{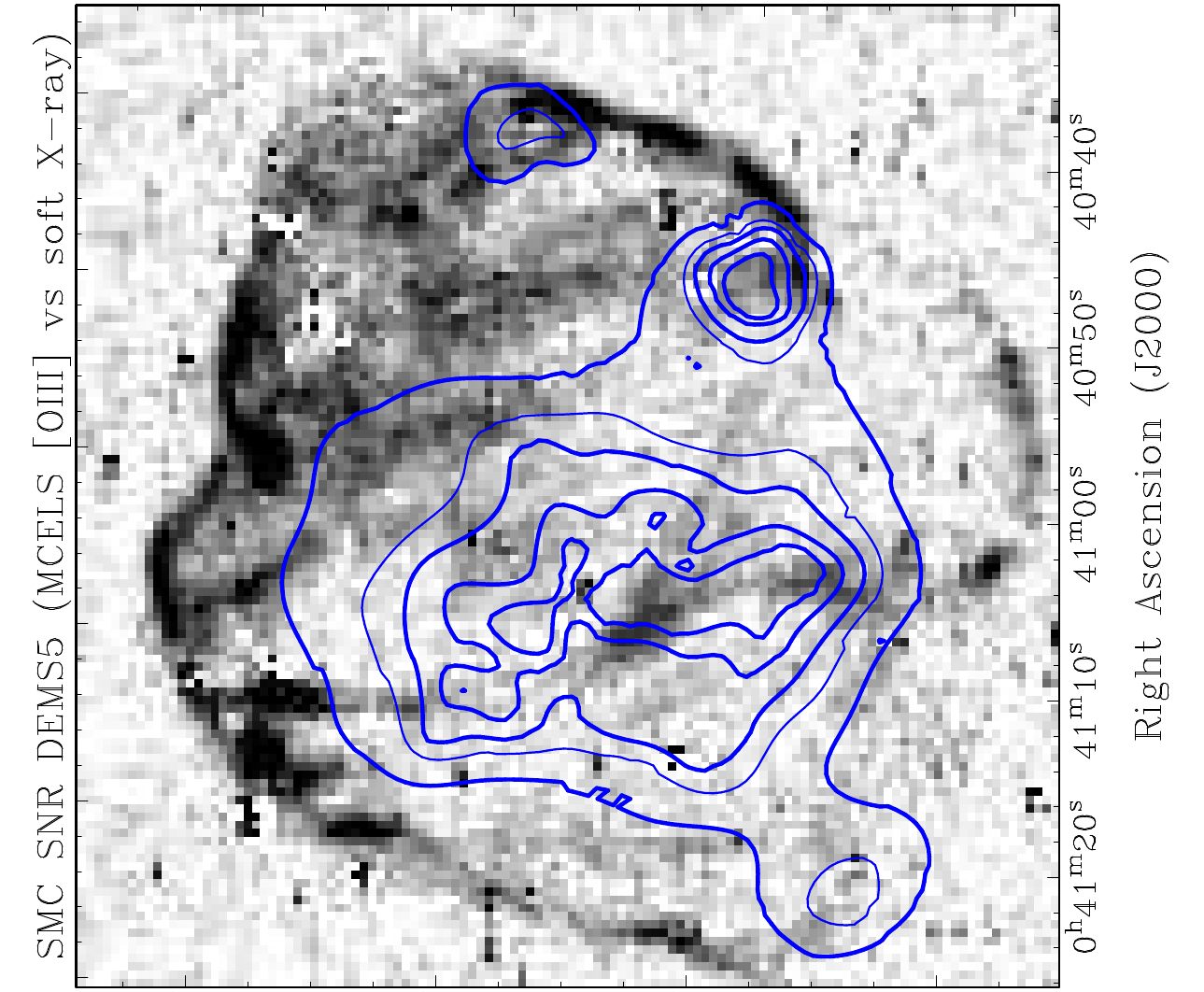}}
    \caption{MCELS images of \dem\ in \Halpha\ (left; overlaid with the hard X-ray (2--4.5\,keV) contours (red)), \SII\ and \OIII\ (middle and right panel; overlaid with the soft band (0.2--1\,keV) X-ray contours (blue)). X-ray contour levels are the same as in Fig.~\ref{fig5}. }  
   \label{fig8}
\end{figure*}

\subsection{Radio Polarisation}
\label{polarisation}
Linear polarisation images of \dem\ at 5500\,MHz were created using the \textit{Q} and \textit{U} Stokes parameters as shown in Fig.~\ref{fig2}. However, no reliable polarisation images could be created at any other frequency, due to the poor signal-to-noise ratio (9000\,MHz) and depolarisation (2100\,MHz) and/or missing short spacings. The fractional polarisation at 5500\,MHz has been evaluated using the standard \textsc{miriad} task \texttt{impol}. The majority of the polarised emission is located at the position of the suggested PWN. Our estimated peak fractional polarisation value is P=~32~$\pm$~7~\,percent while average polarisation is about 23\,percent. Such polarisation that we detected in this area would disfavor extragalactic origin of the object and strongly advocate for the SMC birthplace. Moreover, this is in agreement with our other LMC PWN measurements such as \citet{2012A&A...543A.154H}.

\subsection{Optical Morphology}
The outer borders of \dem\ show that \OIII\ emission dominates in this part of the remnant (Figs.~\ref{fig5}, \ref{fig7} and \ref{fig8}). The X-rays show no oxygen enhancement and the \OIII\ emission clearly delineates the outer blast waves (Fig.~\ref{fig8}; right panel). This indicates that it is likely caused by radiative shocks, since \OIII\ is the strongest coolant of gas at a few 10$^5$\,K. Since \dem\, is probably in a radiative phase it presents very different characteristics than found in much younger SNR like Cas\,A or G292.0+1.8.


\dem\ is some 300\,pc west of the main body of the SMC where the stellar environment and star formation would favour a Type\,Ia scenario \citep{2016A&A...585A.162M}. This is the only known SMC SNR with a lack of recent star formation in the Star Formation History (SFH) map of \citet{2004AJ....127.1531H}. All other SMC SNRs are associated with a recent peak in SFH, although some might be just in projection. However, it is possible that the progenitor of \dem\ was itself a companion in a binary system that got ejected when the primary exploded (i.e. a runaway star), as the linear distance traveled by a runaway star would be $\sim 1000\ (t/10\ {\rm\,Myr})\ (v/100~{\rm km\,s^{-1}}) $\,pc. Therefore, it is possible that if the progenitor of \dem\ was in a binary system with a much more massive (and thus short-lived) companion, it had time to reach its current location, devoid of intense recent star-forming activity.
Since the orientation of the PWN propagation is also away from the main body of the SMC, we consider the possibility that the motion of \dem\ PWN could reflect the peculiar velocity of the progenitor star instead of a kick. However, the runaway velocity is unlikely to be that large, so possibly some compounded velocities (progenitor runaway~+~pulsar kick) are needed.

The central region is dominated by a keV-temperature thermal plasma from the X-rays (Fig.~\ref{fig8}) that runs in the north-south direction of \dem. Diffuse X-ray emission to the left (east) in the interior is likely thermal emission from either shock-heated ISM or ejecta that was overran by the reverse shock, which now should have dissipated for the whole SNR. Therefore, the SNR is probably in a late stage of evolution and the missing or very weak radio synchrotron emission should support that as well as the missing thermal X-ray emission from the shell. The radius differences seem to indicate that the right hand part (west) is further evolved and therefore expanding in a medium of higher density. The missing or very weak X-ray emission from the ejecta should support that too. We also recognise a somewhat smaller circular shell at the west side of \dem\ in \Halpha\ and \SII\ bands. This could be because of a density gradient across the entire SNR. If so, the explosion site might not be at the geometric centre but offset to the west which could result in a somewhat lower putative pulsar kick velocity.

\subsection{Infrared Morphology}
\begin{figure*}
   \includegraphics[scale=0.8,angle=-90,trim=0 0 0 0,clip]{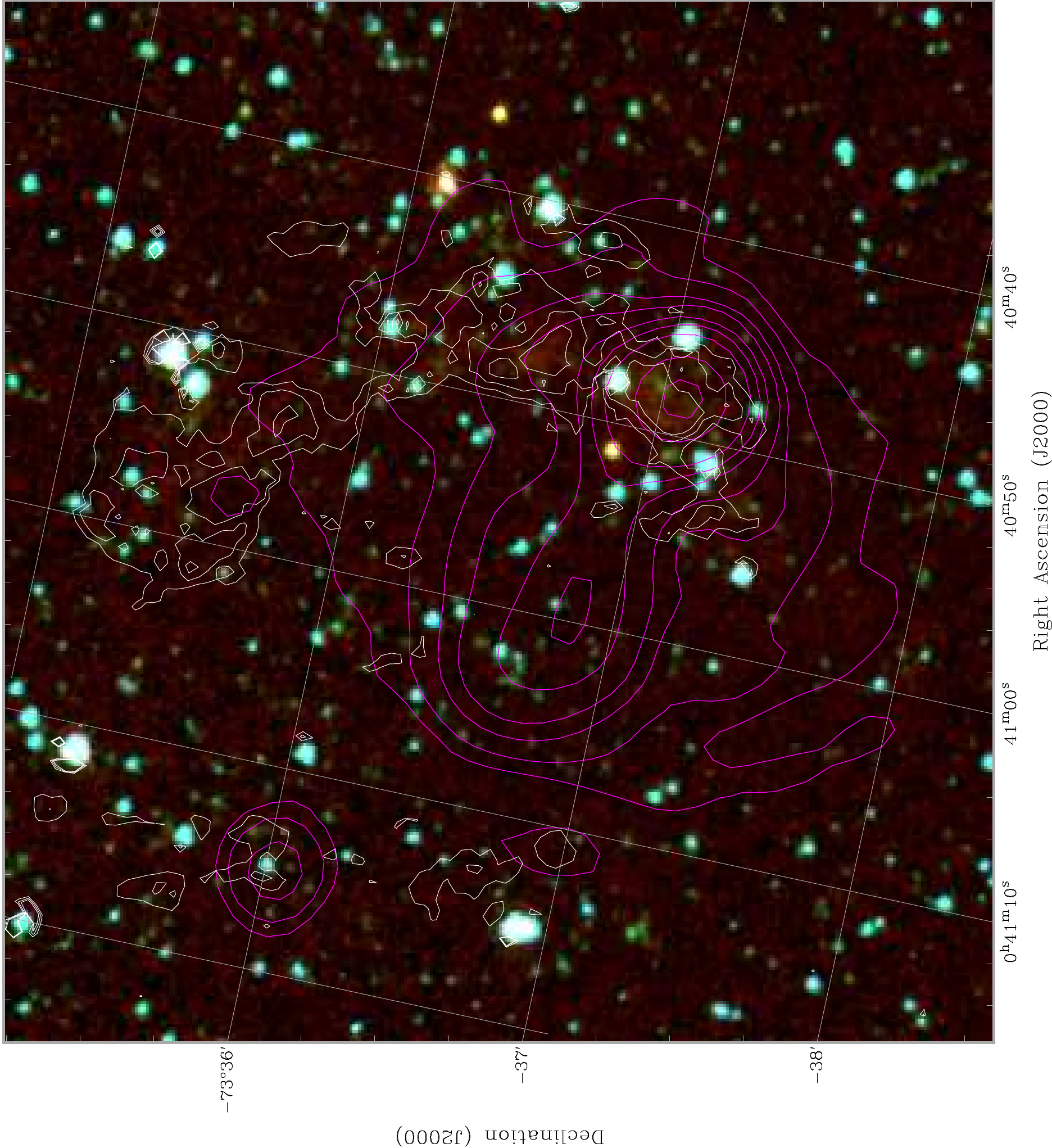}
    \caption{ {\it Spitzer} RGB colour image of \dem\ overlaid with ASKAP (1320\,MHz) contours (magenta) and \SII\ in white. Contours are the same as in Fig.~\ref{fig1}. RGB images correspond to {\it Spitzer} IRAC band-4 (R; 8\,$\mu$m), IRAC band-2 (G; 4.5\,$\mu$m) and IRAC band-1 (B; 3.6\,$\mu$m). Mid-IR emission nicely matches \SII\ (and \Halpha) indicating the presence of shocked gas.  }  
   \label{fig9}
\end{figure*}

We  used the {\it Spitzer} SAGE survey \citep{2011AJ....142..102G} to search for any extended mid-IR emission in the direction of \dem. We found enhanced emission in several bands with 8\,$\mu$m (IRAC band-4) being the most prominent (see Fig.~\ref{fig9}). The IR emission clearly extends towards the north-west  of \dem\ and tightly follows the \SII/\Halpha\ emission in this region just above the radio PWN. Neither the optical nor the IR emission peaks precisely at the putative pulsar position. However, we interpret this displacement as shocked gas trailing the whole PWN \dem\ system. Alternatively, it could come from the SNR shock itself.   

\begin{figure*}
   \includegraphics[width=\textwidth]{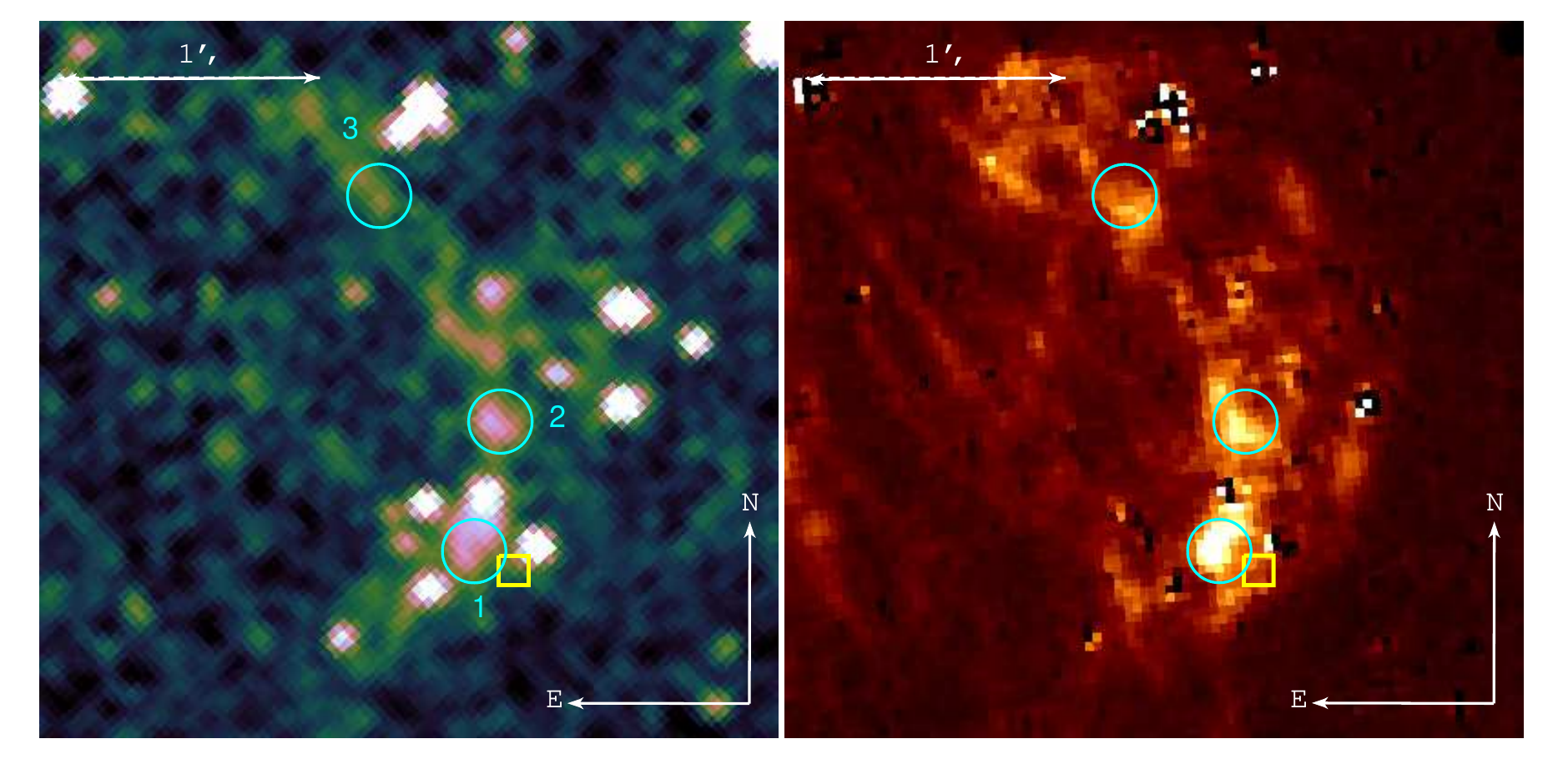}
    \caption{ {\it Spitzer} western shock front of SNR \dem\ as viewed in the IRAC-4 band (8\,$\mu$m; left panel) and the MCELS \SII\ (right panel). Approximate position of the pulsar is marked with the yellow box. Indicated are the three apertures used to measure the mid-IR emission, each is 7.5\,arcsec in radius, centered on the bright emission and avoiding field stars. These mid-IR hotspots are also bright in the radio continuum, indicating that at least some (if not most) of the emission arises from synchrotron.
    }  
   \label{fig10}
\end{figure*}

To ascertain the nature of the mid-IR nebulosity, we measure the integrated flux density in the brightest hotspots, and compare across the IRAC colors to discern the dominant emission components. The field is dense with stars, notably in the IRAC-1 band image, and we carefully choose the apertures to avoid stars while also maximizing the diffuse emission captured in the aperture. Three positions are identified, with apertures chosen to be 7.5\,arcsec in radius, centered on the equatorial J2000 positions: (\#1: 10.203408\D,~--73.617633\D), (\#2: 10.198245\D,~--73.608746\D) and (\#3: 10.225005\D,~--73.594583\D). These positions correspond to the brightest IRAC-4 band nebulosity near the PWN and along the northern extension; see Figs.~\ref{fig9} and ~\ref{fig10}.

Emission is detected in all four IRAC bands for the three positions; the integrated flux density measurements (magnitude and mJy units) for the first and brightest position (marked \#1 in Fig.~\ref{fig10}) are shown in Table~\ref{tab5}. We then compute the IRAC colors ratios: F(4.5)/F(8.0) and F(3.6)/F(5.8), which are used to decode the emission. For the brightest hotspot, the corresponding colors are 0.56 and 0.41, respectively. The other two positions have similar colors: (0.46, 0.37) and (0.44, 0.35) for positions \#2 and \#3, respectively. The errors of the IRAC colour ratios are small ($<$~few\,percent). We now consider the nature of the emission based on these colors. 

\begin{figure}
   \includegraphics[width=\columnwidth]{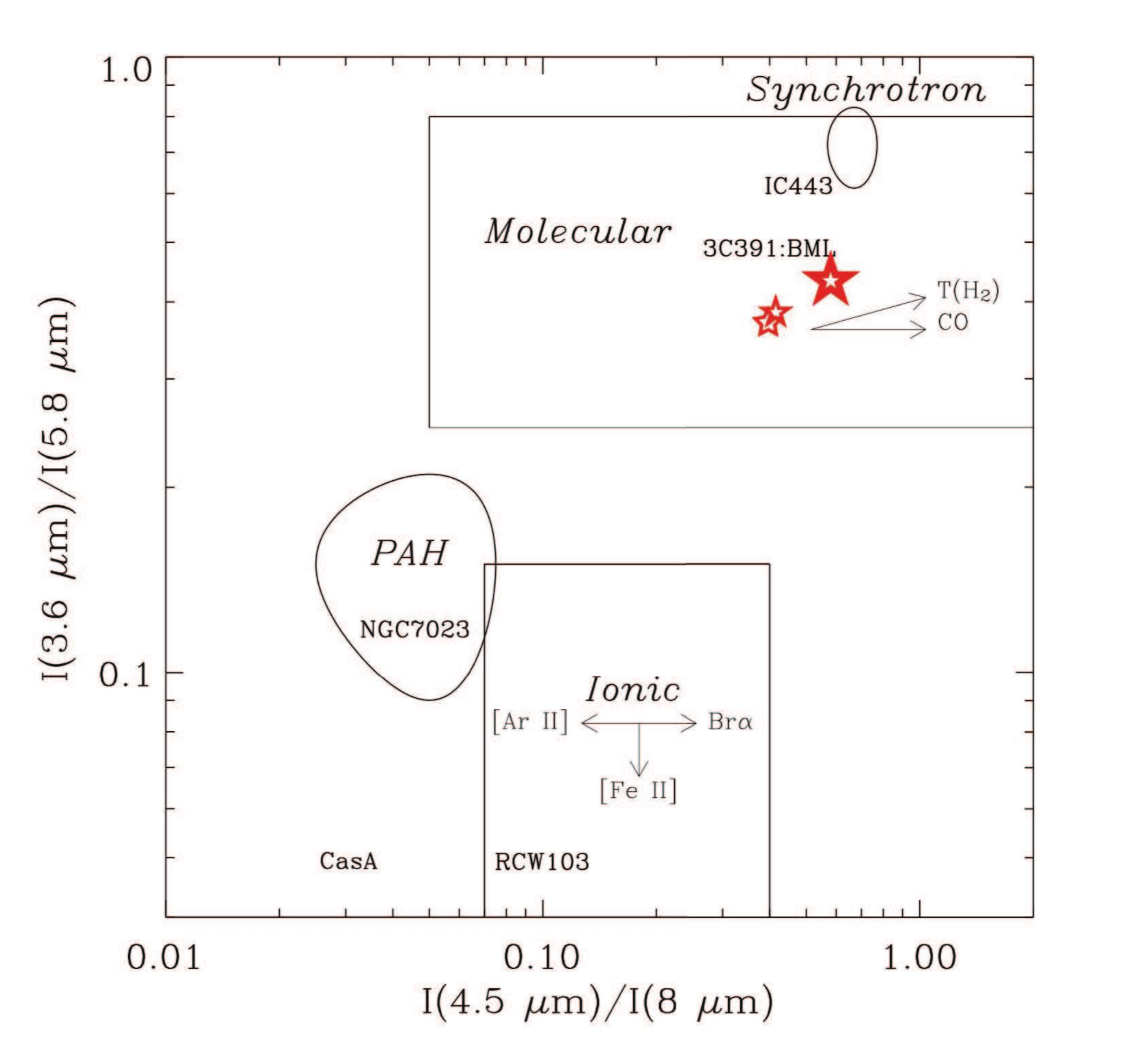}
    \caption{ {\it Spitzer}  IRAC color-color diagram, adapted from \citet{2006AJ....131.1479R}, shows where the dominant emission mechanism appear in color space. Three basic classes are shown: PAH and \HII\ emission from star formation, ionic shock emission and molecular shock emission. Based on the measurements in the three apertures (Figs.~\ref{fig9} and ~\ref{fig10}), the mid-IR emission in \dem\ has colors (indicated by the red stars) that clearly indicate molecular hydrogen emission is dominating the broad-band imaging. }    
   \label{fig11}
\end{figure}

Mid-IR emission arises from atomic, molecular and dust species, excited by star formation and shocks in the ISM. The four broad-bands of IRAC have been effectively used to study SNRs in the Milky Way, including those embedded within dense gas. \citet{2006AJ....131.1479R} modeled the major emission mechanisms and created a set of diagnostics to assess the mid-IR emission associated with SNRs \citep[see their figs.~1 and 2]{2006AJ....131.1479R}. They considered thermal (star formation) and shock excitation, with the latter divided between the diffuse and dense gas ($n~\sim$~$>$~10$^2$\,cm$^{-3}$) ISM.  For example, where ionic line emission dominates, the IRAC-3 and 4 bands tend to be very bright relative to the other bands due to the presence of strong \ArII\ and \FeII\ lines (see their fig.~1). The Galactic SNR~RCW103 is a classic example of IRAC colors influenced by ionic-dominated lines. Alternatively, if the blast wave and subsequent shock passes through dense gas, the dominant coolant is molecular hydrogen, and in conjunction with a molecular CO band head in the IRAC-2 band window, the colors conspire to look very different from the ionic case. The other Galactic SNR~IC443 is an example of a SNR embedded within a molecular cloud. The color-color diagnostic of \citet{2006AJ....131.1479R} is shown in Fig.~\ref{fig11}, where we have plotted the colors of the \dem\ mid-IR nebulosity. 

Based on the IRAC mid-IR emission diagnostic, the \dem\ nebulosity is consistent with shock-excited molecular hydrogen. It has colors that are similar to the notably (molecular cloud) embedded SNRs, such as in Galactic SNRS IC443, W44, G311.5, and G346.6 \citep[see fig.~22]{2006AJ....131.1479R}. The IR emission is spatially coincident with the radio continuum (see Fig.~\ref{fig10}), which may indicate synchrotron contribute to the infrared signal at some low level. Nevertheless, the IRAC colors clearly indicate that molecular emission is the dominant cooling mechanism of the shock front seen in \dem.  

\begin{table*}
\caption{Integrated flux density (7.5~arcsec radius aperture) of the mid-IR emission for the brightest hotspot (\#1 from Fig.~\ref{fig10}).}
\begin{tabular}{l c c c c c c c c c c}
\hline \hline
 Band  &  RA(J2000)& Dec(J2000)  & f     & mag    & dmag  & S/N  \\
       &  (degree)     & (degree)        &(mJy)  &        &       &      \\
 \hline
 IRAC-1 & 10.20294 & --73.61740 & 0.146 & 15.714 & 0.396 & 2.7  \\
 IRAC-2 & 10.20309 & --73.61750 & 0.236 & 14.704 & 0.309 & 3.5  \\
 IRAC-3 & 10.20324 & --73.61760 & 0.357 & 13.769 & 0.272 & 4.0  \\
 IRAC-4 & 10.20380 & --73.61760 & 0.425 & 12.946 & 0.236 & 4.6  \\
   \hline
\end{tabular}\\
\label{tab5}
\end{table*}

While it is not unusual to detect SNRs in the IR bands \citep[see e.g.][]{2006AJ....131.1479R,2007MNRAS.378.1237B,2015ApJ...799...50L} this would be the first bow shock supersonic PWN that potentially shows IR emission compared to a population of 28 described by \citet[see their table~1]{Kargaltsev2017}. However, there are a few cases in our Galaxy and in the LMC of detected IR emission from PWN, e.g.~B0540--69.3~\citep{2008ApJ...687.1054W,2014ApJ...780...50B}, G54.1+0.3 \citep{2010ApJ...710..309T}, and others. Still, none of these bright IR SNRs have bow-shock PWN that moves supersonically. At the same time, the prominent IR emission from~\dem\ clearly following the \SII\ emission which would therefore argue for a more classical SNR origin, with the IR emission associated with the outer SNR shock rather than with the PWN/bow shock. Also, \dem\ is just the second SMC SNR (among a 25 SNR strong population) detected in IR along with SMC SNR~J0106--7205 (a.k.a. IKT\,25) \citep{2007PASJ...59S.455K} indicating that the SNR appearance in IR frequencies are indeed rare. However, the sensitivity of the present generation of IR telescopes and especially at the distance of the SMC might play an important role in the poor detection rate of IR emission from the SMC SNRs.

\subsection{\HI\ Morphology}
\label{hi}
\begin{figure*}
   \includegraphics[width=\textwidth]{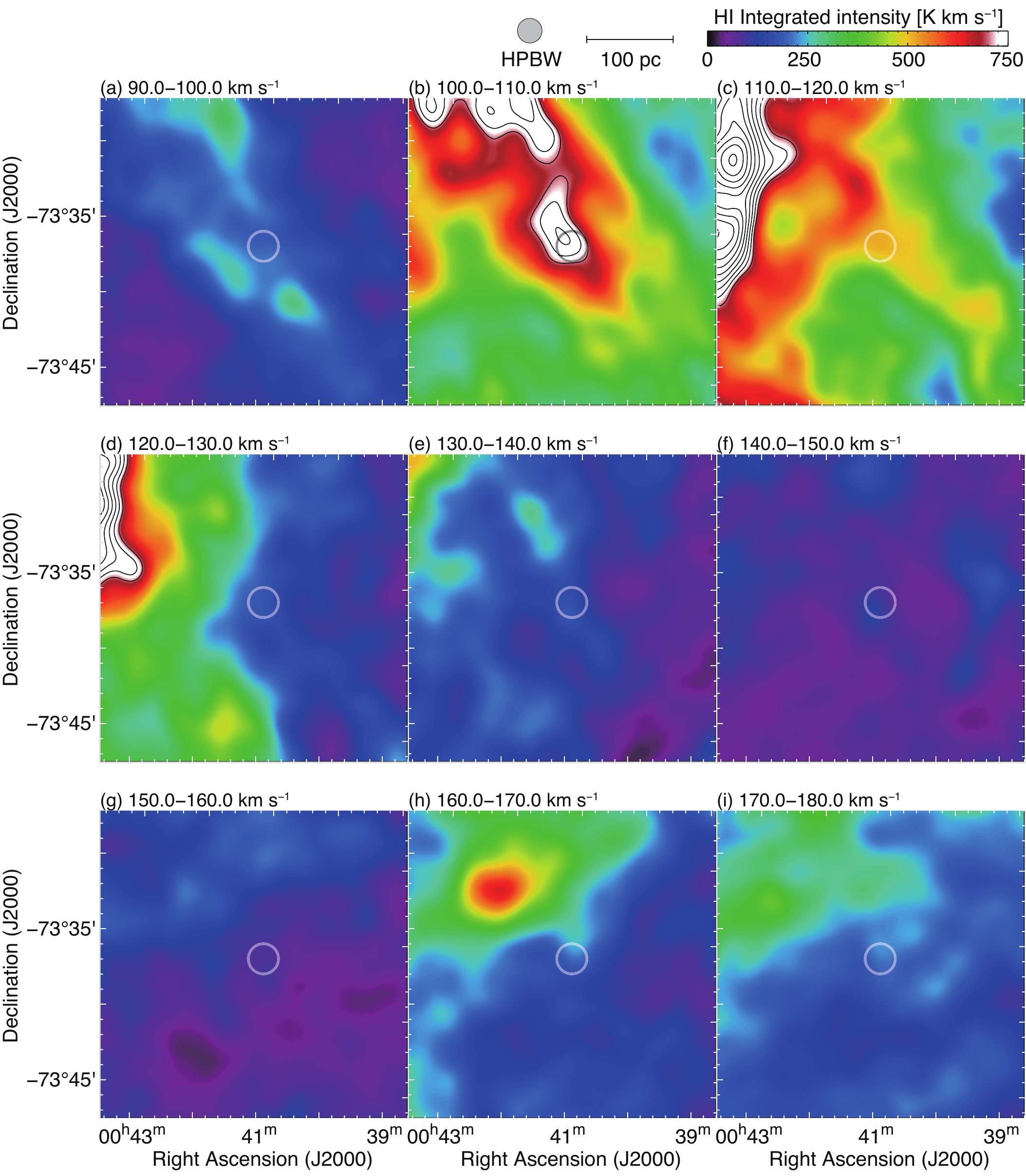}
    \caption{Velocity channel maps of \HI\ toward SNR \dem\ obtained with the ATCA and Parkes telescopes \citep{Stanimiro1999}. Each panel shows \HI\ intensity distribution integrated every 10\,km\,s$^{-1}$ in a velocity range from 90 to 180\,km\,s$^{-1}$. Superposed circles indicate a shell boundary of the SNR. The scale bar and beam size of \HI\ are also shown.}    
   \label{fig12}
\end{figure*}

\begin{figure*}
   	\includegraphics[width=16cm]{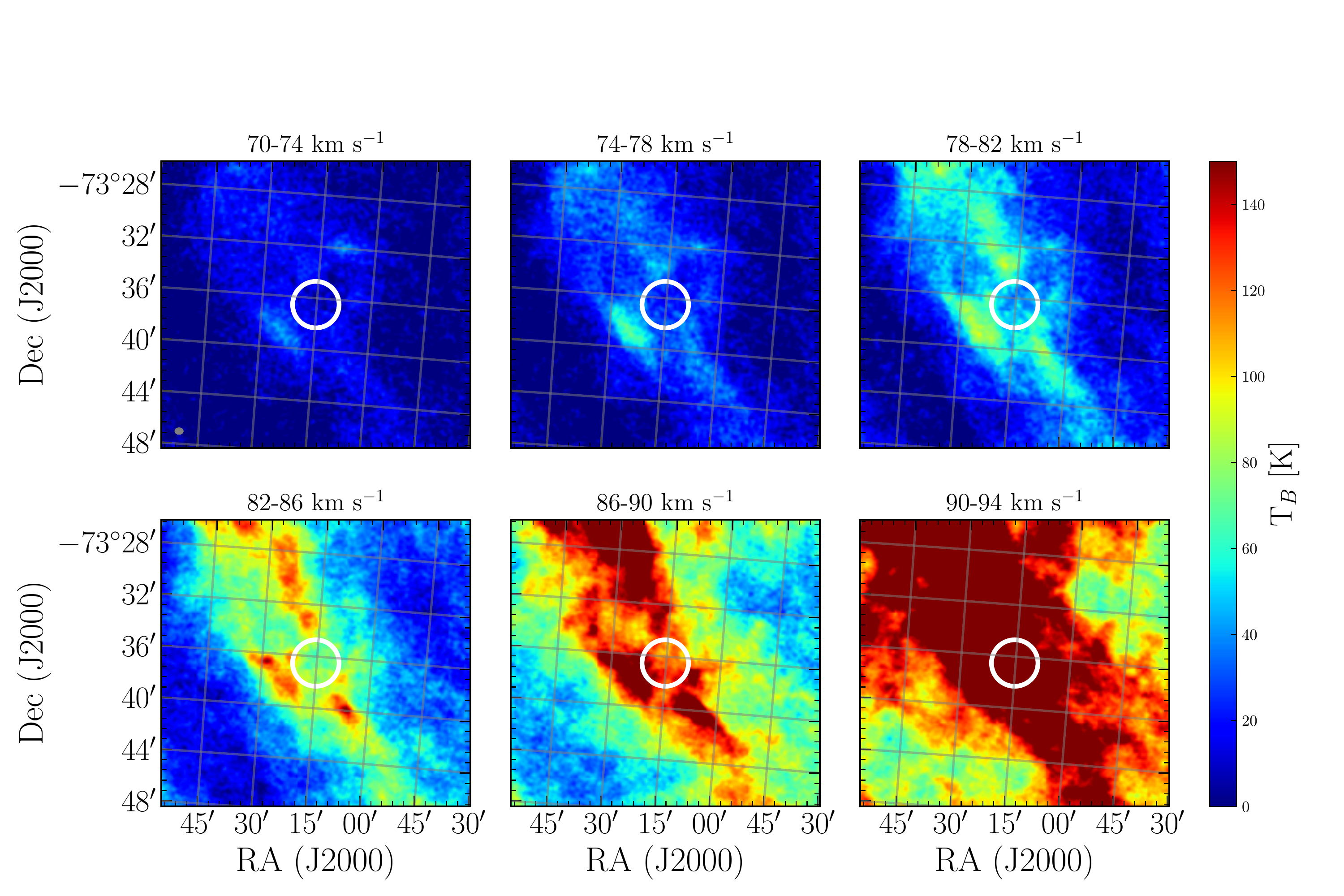}
	\includegraphics[width=16cm]{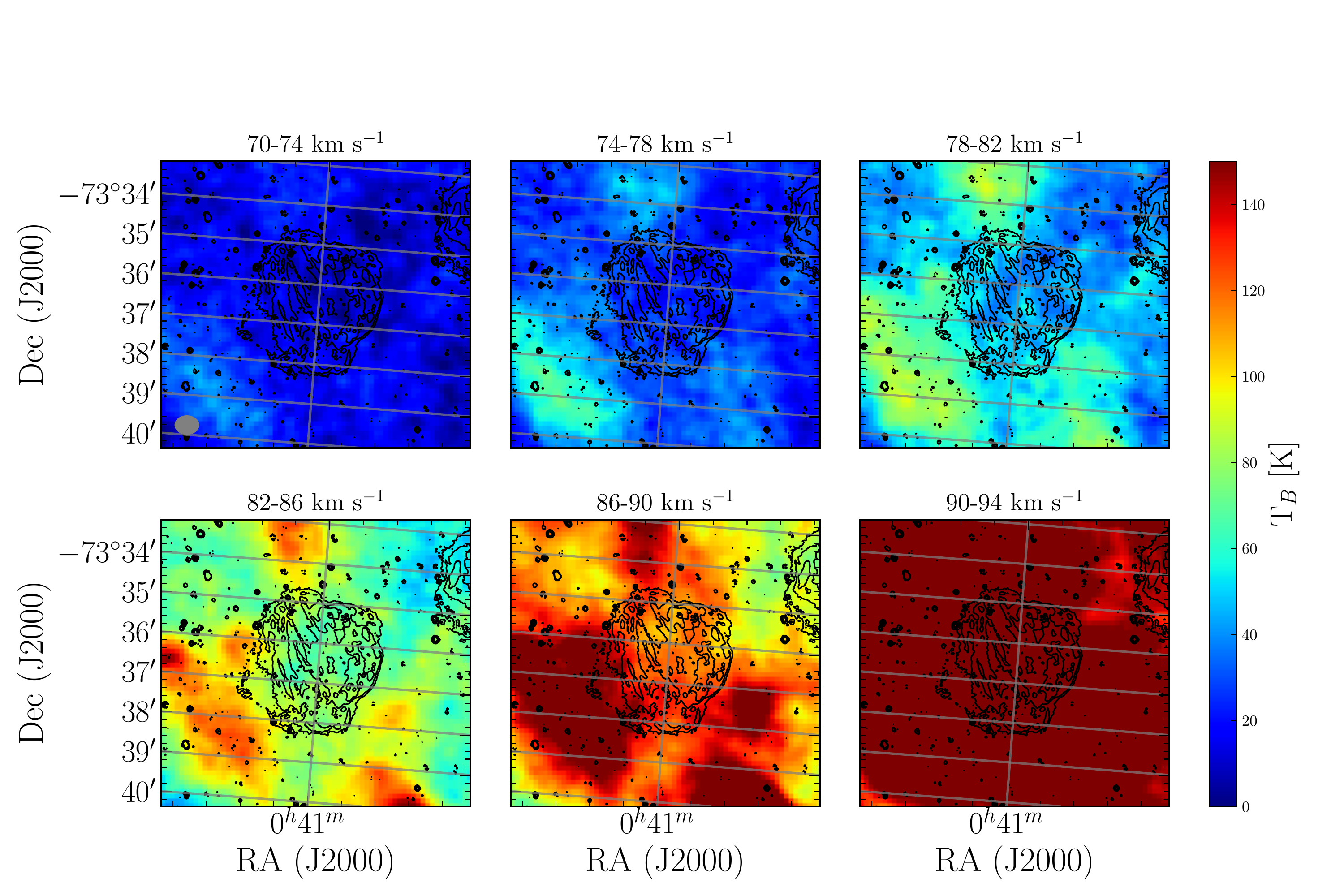}	
   	\caption{Integrated \HI\ moment maps (70--94\,km\,s$^{-1}$) toward SNR \dem\ obtained with the ASKAP and Parkes telescopes \citep{2018NatAs...2..901M}. The maps are integrated over 4\,km\,s$^{-1}$. A colour bar is shown on the right hand side and the beam size is shown at the bottom left corner of the first images. The top panel shows a 22\,arcmin region around SNR \dem. The white circle indicates the size and location of the SNR. The bottom panel shows the same region zoomed into the inner 7\,arcmin and with the H$\alpha$ contours (25, 50, 100, and 200~$\times$~10$^{-17}$\,ergs\,cm$^{-2}$\,s$^{-1}$) overlayed.}
  	\label{fig13}
\end{figure*}

\begin{figure*}
   \includegraphics[width=\textwidth]{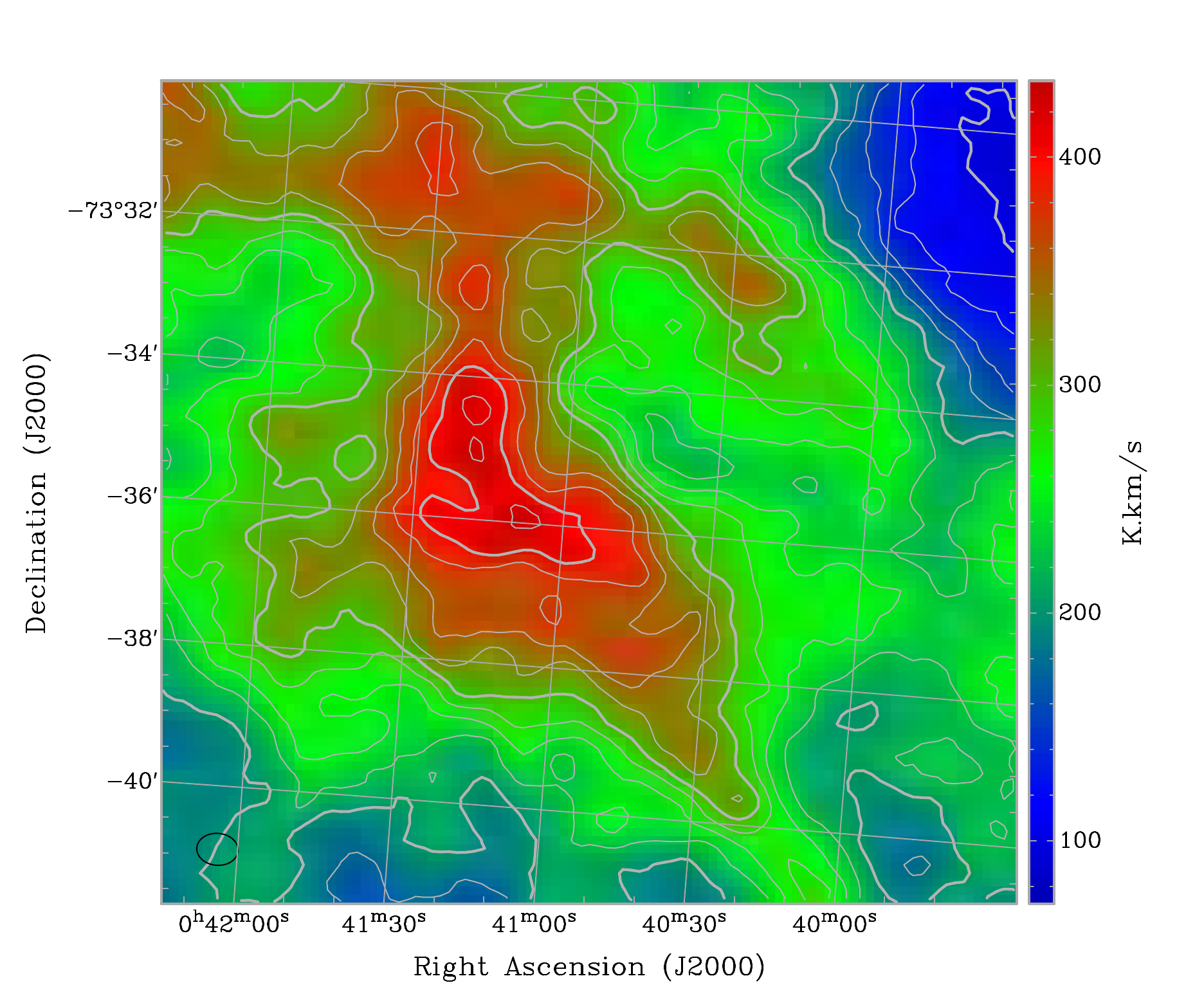}
    \caption{An integrated \HI\ moment map toward SNR \dem\ obtained with the ASKAP and Parkes telescopes \citep{2018NatAs...2..901M}. The map is integrated over two channels (8\,km\,s$^{-1}$) centred at a heliocentric velocity of 100\,km\,s$^{-1}$. The contours are spaced at intervals of 25\,K\,km\,s$^{-1}$ starting at 75\,K\,km\,s$^{-1}$. Bold contours are at intervals of 100\,K\,km\,s$^{-1}$ starting at 100\,K\,km\,s$^{-1}$. A colour bar is shown on the right hand side and the beam size is shown at the bottom left. 
    }    
   \label{fig14}
\end{figure*}

Fig.~\ref{fig12} shows the velocity channel maps of \HI\ obtained with the ATCA and Parkes telescopes \citep{Stanimiro1999}. We find two prominent \HI\ peaks at the radial velocities of $\sim$107\,km\,s$^{-1}$ (hereafter low-velocity cloud) and $\sim$166\,km\,s$^{-1}$ (hereafter high-velocity cloud). The intensity peak of the high-velocity cloud is $\sim$100\,pc away from the SNR (Fig.~\ref{fig12}; bottom middle panel) and unlikely to be associated.  For the low-velocity cloud the intensity peak shows a good spatial correspondence with the position of the SNR (Fig.~\ref{fig12}; top middle panel), and a possible cavity-like structure is seen (top left panel). Recent data taken with the ASKAP telescope of the SMC \citep{2018NatAs...2..901M} shows this cavity in more detail in Fig.~\ref{fig13}. The ASKAP data has three times better spatial resolution (35~$\times$~27\,arcsec) compared to ATCA data and an RMS of 0.7\,K per a 3.9\,km\,s$^{-1}$ wide spectral channel. Fig.~\ref{fig13} shows a filament like \HI\ structure that has a cavity corresponding to the location and size of the H$\alpha$ emission of SNR \dem. A moment map in this velocity range of 96--104\,km\,s$^{-1}$ of the ASKAP data is shown in Fig.~\ref{fig14}.

However, the radius of this cavity ($\sim$70\,pc) is over twice as large as the SNR, so may not be directly associated, or may represent the combined energy input from the progenitor and previous generations of nearby stars and SNe. Possible line-splitting is present. If interpreted as shell expansion, the expansion velocity is $\sim$8\,km\,s$^{-1}$, which leads to a dynamical age of 5\,Myr, an average mechanical luminosity of $\sim$400\,L$_{\sun}$ and a total energy of $2\times 10^{50}$\,erg, assuming an ambient density of 2\,cm$^{-3}$ \citep{1997MNRAS.289..225S}. These values indicate that an association with the SNR is plausible, particularly if the SNR has re-accelerated a shell or pre-existing structure created by past events.


\section{Discussion}
\label{Sect:discussion}
We explore two possible scenarios for the nature of this object. Namely:
\begin{enumerate}
\item A PWN associated with \dem:  
This is supported by (a) the detection of a radio point source coincident with the hard X-ray point source (PWN/pulsar composite), (b) an X-ray spectrum consistent with a power law with photon index typical as seen from PWN/pulsar spectra and the stability of the X-ray flux, (c) the radio spectral indices and (d) shocked IR emission. In the case of PWNe, the extent of the radio emission is generally greater than that in X-rays because of the longer cooling time of radio-emitting electrons. 
The radio emission may therefore trace the path of the PWN from the centre of the SNR as indicated by the current radio data of \dem\ at 2100\,MHz (Fig.~\ref{fig5}). The detection of the cometary morphology in radio indicates a large kick velocity of the pulsar at birth. The estimated distance between the point-like source and the centre of the SNR indicates that the transverse velocity of the putative pulsar could be as high as $\sim$2000\,km\,s$^{-1}$ with an age of 10--15\,kyr. The lower and somewhat more realistic limits would put this SNR/PWN system into $\sim$28\,kyr age and kick-off velocity of 700--800\,km\,s$^{-1}$. Determination of the putative pulsar velocity is important to determine the birth kick distribution of the NS which can further constrain supernova explosion models and kick mechanisms \cite[e.g.][]{Ng2007}. Finding such a source in the SMC is particularly exciting as there might be differences in the pulsar birth, explosion and kick properties in the low-metallicity environment of SMC.

To some extent \dem\ resembles similar morphological appearance like two well-studied LMC SNRs -- MCSNR\,J0508--6902 \citep{2014MNRAS.439.1110B} and MCSNR\,J0527--7104 \citep{2016A&A...586A...4K} as well as, for example, Galactic SNR MSH~11-61A \citep{2005SerAJ.170...47F} or G332.5--5.6 \citep{2007MNRAS.381..377S}. While two LMC SNRs are classified as a likely Type~Ia, they all have apparent `mixed-morphology' appearance where SNR looks quite different at different wavebands. Mixed-morphology SNRs have shell-like radio morphologies combined with center-filled X-ray morphologies. The central X-ray morphology is produced by thermal emission rather than a central PWN \citep{1998ApJ...503L.167R,2017ApJ...839...59P}. What makes \dem\ different from other so called `mixed-morphology' SNRs is its bow-shock PWN.


\item An AGN in the background of the SMC: It is possible that the hard X-ray point source is not associated with the SNR but is instead a background AGN coincident with the SMC. The diffuse radio emission would in this case be due to a radio lobe or jets. A greater absorption component would then be expected to account for the line of sight through the SMC, although this may be sometimes balanced by a soft excess component \citep{Turner1989}. As seen in Section~\ref{xray}, the absorption at the source position through the depth of the SMC indicates that the source is most likely located in the SMC. We also searched for a possible AGN association of the source using the ALLWISE criterion in the mid-infrared and from the existing quasar catalogs. We did not find any AGN within 3$\sigma$ of the source position \citep{2018A&A...612A..87M}.
\end{enumerate}

\section{Conclusions and Future Studies}
\label{Sect:conclusion}

We have presented results of a newly discovered pulsar-powered nebula associated with SNR \dem\ in the SMC. The object exhibits a cometary morphology which suggests a pulsar leading the PWN and traveling supersonically through the ambient medium. This is the first detected extragalactic source to exhibit this type of morphology. The distance of this object ($\sim$21.2\,pc) from the centre of \dem\ SNR would require a kick velocity as high as $\sim$2000\,km\,s$^{-1}$ and as low as $\sim$700\,km\,s$^{-1}$ for the pulsar at birth. The SNR/PWN/Pulsar system exhibits a typically flat SED of --0.29~$\pm$~0.01 inline with most other PWNe. We found no convincing DMs candidates (up to 1000\,cm$^{-3}$\,pc) with a S/N greater than 8. We also detect radio polarisation in the locale of the emission from this object with a peak (fractional) value at 5500\,MHz of P=~32~$\pm$~7\,percent and average of $\sim$23\,percent which strongly indicates non-thermal emission. At IR frequencies, we detect associated shock emission that indicates that the \dem\ SNR is still expanding. Finally, our \HI\ velocity channel maps show possible interaction with the low-velocity cloud at $\sim$107\,km\,s$^{-1}$.

Future follow-up multi-frequency observations of this object with a high spatial resolution in X-rays will be able to accurately measure the extent of the hard X-ray point source within \dem, and resolve the compact object from the diffuse nebula. We will also search for a radio pulsar at this position. Another objective will be to accurately measure the position of the compact object within \dem, which will help to better constrain the motion of the pulsar within the SNR and identify its counterpart at other wavelengths. 

Finally, the study of PWNe at TeV gamma-ray energies (where the inverse-Compton process provides the gamma-ray photons) allows important constraints on the magnetic field and transport properties of the PWN electrons \cite[e.g.][]{1997MNRAS.291..162A}. The now well-established X-ray and TeV gamma-ray emission correlation in PWNe \citep{2009ApJ...694...12M,2013arXiv1305.2552K, 2018A&A...612A...3H} can provide a prediction of the potential TeV flux. For the pulsar spin-down power and characteristic age of \dem, the TeV flux could be a factor 5 to 10 times higher than its X-ray flux (3.6~$\times$~10$^{-14}$\,\ergcm\~). Such a TeV flux level is within the reach of the forthcoming \v{C}erenkov Telescope Array (CTA) \citep{cta1,cta2}. A similar X-ray to TeV flux scaling may apply to the other PWN in the SMC, IKT\,16 \citep{2015A&A...584A..41M}, bringing it within the reach of the CTA.

\section*{Acknowledgements}
The Australian Compact Array and the Australian SKA Pathfinder (ASKAP) are part of the Australian Telescope which is funded by the Commonwealth of Australia for operation as National Facility managed by CSIRO. This paper includes archived data obtained through the Australia Telescope Online Archive (http://atoa.atnf.csiro.au). We used the \textsc{karma} and \textsc{miriad} software packages developed by the ATNF. Operation of ASKAP is funded by the Australian Government with support from the National Collaborative Research Infrastructure Strategy. ASKAP uses the resources of the Pawsey Supercomputing Centre. Establishment of ASKAP, the Murchison Radio-astronomy Observatory and the Pawsey Supercomputing Centre are initiatives of the Australian Government, with support from the Government of Western Australia and the Science and Industry Endowment Fund. We acknowledge the Wajarri Yamatji people as the traditional owners of the Observatory site.
M.S.\ acknowledges support by the Deutsche Forschungsgemeinschaft (DFG) through the Heisenberg research grant SA 2131/4--1 and the Heisenberg professor grant SA 2131/5--1.
This work is part of project No.~176005, `Emission nebulae: structure and evolution', supported by the Ministry of Education, Science, and Technological Development of the Republic of Serbia. 
Parts of this research were conducted with support of Australian Research Council Centre of Excellence for All Sky Astrophysics in 3 Dimensions (ASTRO~3D), through project number CE170100013.   
The authors would like to thank the anonymous referee for a constructive report and useful comments. 







\bsp	
\label{lastpage}


\begin{thebibliography}{}
\makeatletter
\relax
\def\mn@urlcharsother{\let\do\@makeother \do\$\do\&\do\#\do\^\do\_\do\%\do\~}
\def\mn@doi{\begingroup\mn@urlcharsother \@ifnextchar [ {\mn@doi@}
  {\mn@doi@[]}}
\def\mn@doi@[#1]#2{\def\@tempa{#1}\ifx\@tempa\@empty \href
  {http://dx.doi.org/#2} {doi:#2}\else \href {http://dx.doi.org/#2} {#1}\fi
  \endgroup}
\def\mn@eprint#1#2{\mn@eprint@#1:#2::\@nil}
\def\mn@eprint@arXiv#1{\href {http://arxiv.org/abs/#1} {{\tt arXiv:#1}}}
\def\mn@eprint@dblp#1{\href {http://dblp.uni-trier.de/rec/bibtex/#1.xml}
  {dblp:#1}}
\def\mn@eprint@#1:#2:#3:#4\@nil{\def\@tempa {#1}\def\@tempb {#2}\def\@tempc
  {#3}\ifx \@tempc \@empty \let \@tempc \@tempb \let \@tempb \@tempa \fi \ifx
  \@tempb \@empty \def\@tempb {arXiv}\fi \@ifundefined
  {mn@eprint@\@tempb}{\@tempb:\@tempc}{\expandafter \expandafter \csname
  mn@eprint@\@tempb\endcsname \expandafter{\@tempc}}}

\bibitem[\protect\citeauthoryear{{Aharonian}, {Atoyan}  \&
  {Kifune}}{{Aharonian} et~al.}{1997}]{1997MNRAS.291..162A}
{Aharonian} F.~A.,  {Atoyan} A.~M.,   {Kifune} T.,  1997, \mn@doi [\mnras]
  {10.1093/mnras/291.1.162}, \href
  {https://ui.adsabs.harvard.edu/#abs/1997MNRAS.291..162A} {291, 162}

\bibitem[\protect\citeauthoryear{{Arnaud}}{{Arnaud}}{1996}]{1996ASPC..101...17A}
{Arnaud} K.~A.,  1996, in {Jacoby} G.~H.,  {Barnes} J.,  eds,  Astronomical
  Society of the Pacific Conference Series Vol. 101, Astronomical Data Analysis
  Software and Systems V. p.~17

\bibitem[\protect\citeauthoryear{{Barkov}, {Lyutikov}  \&
  {Khangulyan}}{{Barkov} et~al.}{2019}]{2019MNRAS.484.4760B}
{Barkov} M.~V.,  {Lyutikov} M.,   {Khangulyan} D.,  2019, \mn@doi [\mnras]
  {10.1093/mnras/stz213}, \href
  {https://ui.adsabs.harvard.edu/\#abs/2019MNRAS.484.4760B} {484, 4760}

\bibitem[\protect\citeauthoryear{{Besla}, {Kallivayalil}, {Hernquist},
  {Robertson}, {Cox}, {van der Marel}  \& {Alcock}}{{Besla}
  et~al.}{2007}]{2007ApJ...668..949B}
{Besla} G.,  {Kallivayalil} N.,  {Hernquist} L.,  {Robertson} B.,  {Cox} T.~J.,
   {van der Marel} R.~P.,   {Alcock} C.,  2007, \mn@doi [\apj]
  {10.1086/521385}, \href {http://adsabs.harvard.edu/abs/2007ApJ...668..949B}
  {668, 949}

\bibitem[\protect\citeauthoryear{{Boji{\v c}i{\'c}}, {Filipovi{\'c}}, {Parker},
  {Payne}, {Jones}, {Reid}, {Kawamura}  \& {Fukui}}{{Boji{\v c}i{\'c}}
  et~al.}{2007}]{2007MNRAS.378.1237B}
{Boji{\v c}i{\'c}} I.~S.,  {Filipovi{\'c}} M.~D.,  {Parker} Q.~A.,  {Payne}
  J.~L.,  {Jones} P.~A.,  {Reid} W.,  {Kawamura} A.,   {Fukui} Y.,  2007,
  \mn@doi [\mnras] {10.1111/j.1365-2966.2007.11784.x}, \href
  {http://adsabs.harvard.edu/abs/2007MNRAS.378.1237B} {378, 1237}

\bibitem[\protect\citeauthoryear{{Bozzetto} et~al.,}{{Bozzetto}
  et~al.}{2012}]{2012MNRAS.420.2588B}
{Bozzetto} L.~M.,  et~al., 2012, \mn@doi [\mnras]
  {10.1111/j.1365-2966.2011.20231.x}, \href
  {http://adsabs.harvard.edu/abs/2012MNRAS.420.2588B} {420, 2588}

\bibitem[\protect\citeauthoryear{{Bozzetto} et~al.,}{{Bozzetto}
  et~al.}{2014}]{2014MNRAS.439.1110B}
{Bozzetto} L.~M.,  et~al., 2014, \mn@doi [\mnras] {10.1093/mnras/stu051}, \href
  {http://adsabs.harvard.edu/abs/2014MNRAS.439.1110B} {439, 1110}

\bibitem[\protect\citeauthoryear{{Bozzetto} et~al.,}{{Bozzetto}
  et~al.}{2017}]{2017ApJS..230....2B}
{Bozzetto} L.~M.,  et~al., 2017, \mn@doi [\apjs] {10.3847/1538-4365/aa653c},
  \href {http://adsabs.harvard.edu/abs/2017ApJS..230....2B} {230, 2}

\bibitem[\protect\citeauthoryear{{Brantseg}, {McEntaffer}, {Bozzetto},
  {Filipovic}  \& {Grieves}}{{Brantseg} et~al.}{2014}]{2014ApJ...780...50B}
{Brantseg} T.,  {McEntaffer} R.~L.,  {Bozzetto} L.~M.,  {Filipovic} M.,
  {Grieves} N.,  2014, \mn@doi [\apj] {10.1088/0004-637X/780/1/50}, \href
  {http://adsabs.harvard.edu/abs/2014ApJ...780...50B} {780, 50}

\bibitem[\protect\citeauthoryear{{Cherenkov Telescope Array Consortium}
  et~al.,}{{Cherenkov Telescope Array Consortium} et~al.}{2019}]{cta2}
{Cherenkov Telescope Array Consortium} T.,  et~al., 2019, Science with the
  Cherenkov Telescope Array.
WORLD SCIENTIFIC (\mn@eprint {}
  {https://www.worldscientific.com/doi/pdf/10.1142/10986}),
  \mn@doi{10.1142/10986}, \url
  {https://www.worldscientific.com/doi/abs/10.1142/10986}

\bibitem[\protect\citeauthoryear{{Cioffi}, {McKee}  \& {Bertschinger}}{{Cioffi}
  et~al.}{1988}]{1988ApJ...334..252C}
{Cioffi} D.~F.,  {McKee} C.~F.,   {Bertschinger} E.,  1988, \mn@doi [\apj]
  {10.1086/166834}, \href
  {https://ui.adsabs.harvard.edu/#abs/1988ApJ...334..252C} {334, 252}

\bibitem[\protect\citeauthoryear{{Collier}, {Norris}, {Filipovi{\'c}}  \&
  {Tothill}}{{Collier} et~al.}{2016}]{2016AN....337...36C}
{Collier} J.~D.,  {Norris} R.~P.,  {Filipovi{\'c}} M.~D.,   {Tothill} N.~F.~H.,
   2016, \mn@doi [Astronomische Nachrichten] {10.1002/asna.201512261}, \href
  {https://ui.adsabs.harvard.edu/#abs/2016AN....337...36C} {337, 36}

\bibitem[\protect\citeauthoryear{{Crawford}, {Filipovic}, {de Horta}, {Wong},
  {Tothill}, {Draskovic}, {Collier}  \& {Galvin}}{{Crawford}
  et~al.}{2011}]{2011SerAJ.183...95C}
{Crawford} E.~J.,  {Filipovic} M.~D.,  {de Horta} A.~Y.,  {Wong} G.~F.,
  {Tothill} N.~F.~H.,  {Draskovic} D.,  {Collier} J.~D.,   {Galvin} T.~J.,
  2011, \mn@doi [Serbian Astronomical Journal] {10.2298/SAJ1183095C}, \href
  {http://adsabs.harvard.edu/abs/2011SerAJ.183...95C} {183, 95}

\bibitem[\protect\citeauthoryear{{Crawford}, {Filipovi{\'c}}, {McEntaffer},
  {Brantseg}, {Heitritter}, {Roper}, {Haberl}  \& {Uro{\v
  s}evi{\'c}}}{{Crawford} et~al.}{2014}]{2014AJ....148...99C}
{Crawford} E.~J.,  {Filipovi{\'c}} M.~D.,  {McEntaffer} R.~L.,  {Brantseg} T.,
  {Heitritter} K.,  {Roper} Q.,  {Haberl} F.,   {Uro{\v s}evi{\'c}} D.,  2014,
  \mn@doi [\aj] {10.1088/0004-6256/148/5/99}, \href
  {http://adsabs.harvard.edu/abs/2014AJ....148...99C} {148, 99}

\bibitem[\protect\citeauthoryear{{Dai}, {Johnston}  \& {Hobbs}}{{Dai}
  et~al.}{2017}]{2017MNRAS.472.1458D}
{Dai} S.,  {Johnston} S.,   {Hobbs} G.,  2017, \mn@doi [\mnras]
  {10.1093/mnras/stx2033}, \href
  {https://ui.adsabs.harvard.edu/#abs/2017MNRAS.472.1458D} {472, 1458}

\bibitem[\protect\citeauthoryear{{Filipovic}, {Jones}, {White}, {Haynes},
  {Klein}  \& {Wielebinski}}{{Filipovic} et~al.}{1997}]{1997A&AS..121..321F}
{Filipovic} M.~D.,  {Jones} P.~A.,  {White} G.~L.,  {Haynes} R.~F.,  {Klein}
  U.,   {Wielebinski} R.,  1997, \mn@doi [\aaps] {10.1051/aas:1997317}, \href
  {http://adsabs.harvard.edu/abs/1997A%26AS..121..321F} {121, 321}

\bibitem[\protect\citeauthoryear{{Filipovic}, {Haynes}, {White}  \&
  {Jones}}{{Filipovic} et~al.}{1998}]{1998A&AS..130..421F}
{Filipovic} M.~D.,  {Haynes} R.~F.,  {White} G.~L.,   {Jones} P.~A.,  1998,
  \mn@doi [\aaps] {10.1051/aas:1998417}, \href
  {http://adsabs.harvard.edu/abs/1998A%26AS..130..421F} {130, 421}

\bibitem[\protect\citeauthoryear{{Filipovi{\'c}}, {Bohlsen}, {Reid},
  {Staveley-Smith}, {Jones}, {Nohejl}  \& {Goldstein}}{{Filipovi{\'c}}
  et~al.}{2002}]{2002MNRAS.335.1085F}
{Filipovi{\'c}} M.~D.,  {Bohlsen} T.,  {Reid} W.,  {Staveley-Smith} L.,
  {Jones} P.~A.,  {Nohejl} K.,   {Goldstein} G.,  2002, \mn@doi [\mnras]
  {10.1046/j.1365-8711.2002.05702.x}, \href
  {http://adsabs.harvard.edu/abs/2002MNRAS.335.1085F} {335, 1085}

\bibitem[\protect\citeauthoryear{{Filipovic}, {Payne}  \& {Jones}}{{Filipovic}
  et~al.}{2005a}]{2005SerAJ.170...47F}
{Filipovic} M.~D.,  {Payne} J.~L.,   {Jones} P.~A.,  2005a, \mn@doi [Serbian
  Astronomical Journal] {10.2298/SAJ0570047F}, \href
  {http://adsabs.harvard.edu/abs/2005SerAJ.170...47F} {170, 47}

\bibitem[\protect\citeauthoryear{{Filipovi{\'c}}, {Payne}, {Reid}, {Danforth},
  {Staveley-Smith}, {Jones}  \& {White}}{{Filipovi{\'c}}
  et~al.}{2005b}]{2005MNRAS.364..217F}
{Filipovi{\'c}} M.~D.,  {Payne} J.~L.,  {Reid} W.,  {Danforth} C.~W.,
  {Staveley-Smith} L.,  {Jones} P.~A.,   {White} G.~L.,  2005b, \mn@doi
  [\mnras] {10.1111/j.1365-2966.2005.09554.x}, \href
  {http://adsabs.harvard.edu/abs/2005MNRAS.364..217F} {364, 217}

\bibitem[\protect\citeauthoryear{{Filipovi{\'c}} et~al.,}{{Filipovi{\'c}}
  et~al.}{2008}]{2008A&A...485...63F}
{Filipovi{\'c}} M.~D.,  et~al., 2008, \mn@doi [\aap]
  {10.1051/0004-6361:200809642}, \href
  {http://adsabs.harvard.edu/abs/2008A%26A...485...63F} {485, 63}

\bibitem[\protect\citeauthoryear{{For} et~al.,}{{For}
  et~al.}{2018}]{2018MNRAS.480.2743F}
{For} B.~Q.,  et~al., 2018, \mn@doi [\mnras] {10.1093/mnras/sty1960}, \href
  {https://ui.adsabs.harvard.edu/#abs/2018MNRAS.480.2743F} {480, 2743}

\bibitem[\protect\citeauthoryear{{Gaensler}, {van der Swaluw}, {Camilo},
  {Kaspi}, {Baganoff}, {Yusef-Zadeh}  \& {Manchester}}{{Gaensler}
  et~al.}{2004}]{2004ApJ...616..383G}
{Gaensler} B.~M.,  {van der Swaluw} E.,  {Camilo} F.,  {Kaspi} V.~M.,
  {Baganoff} F.~K.,  {Yusef-Zadeh} F.,   {Manchester} R.~N.,  2004, \mn@doi
  [\apj] {10.1086/424906}, \href
  {http://adsabs.harvard.edu/abs/2004ApJ...616..383G} {616, 383}

\bibitem[\protect\citeauthoryear{{Gooch}}{{Gooch}}{1995}]{1995ASPC...77..144G}
{Gooch} R.,  1995, in {Shaw} R.~A.,  {Payne} H.~E.,   {Hayes} J.~J.~E.,  eds,
  Astronomical Society of the Pacific Conference Series Vol. 77, Astronomical
  Data Analysis Software and Systems IV. p.~144

\bibitem[\protect\citeauthoryear{{Gordon} et~al.,}{{Gordon}
  et~al.}{2011}]{2011AJ....142..102G}
{Gordon} K.~D.,  et~al., 2011, \mn@doi [\aj] {10.1088/0004-6256/142/4/102},
  \href {http://adsabs.harvard.edu/abs/2011AJ....142..102G} {142, 102}

\bibitem[\protect\citeauthoryear{{H.~E.~S.~S. Collaboration}
  et~al.,}{{H.~E.~S.~S. Collaboration} et~al.}{2018}]{2018A&A...612A...3H}
{H.~E.~S.~S. Collaboration} et~al., 2018, \mn@doi [\aap]
  {10.1051/0004-6361/201732125}, \href
  {https://ui.adsabs.harvard.edu/#abs/2018A&A...612A...3H} {612, A3}

\bibitem[\protect\citeauthoryear{{Haberl} \& {Sturm}}{{Haberl} \&
  {Sturm}}{2016}]{Haberl2016}
{Haberl} F.,  {Sturm} R.,  2016, \mn@doi [\aap] {10.1051/0004-6361/201527326},
  \href {http://adsabs.harvard.edu/abs/2016A%26A...586A..81H} {586, A81}

\bibitem[\protect\citeauthoryear{{Haberl}, {Sturm}, {Filipovi{\'c}}, {Pietsch}
  \& {Crawford}}{{Haberl} et~al.}{2012a}]{2012A&A...537L...1H}
{Haberl} F.,  {Sturm} R.,  {Filipovi{\'c}} M.~D.,  {Pietsch} W.,   {Crawford}
  E.~J.,  2012a, \mn@doi [\aap] {10.1051/0004-6361/201118369}, \href
  {http://adsabs.harvard.edu/abs/2012A%26A...537L...1H} {537, L1}

\bibitem[\protect\citeauthoryear{{Haberl} et~al.,}{{Haberl}
  et~al.}{2012b}]{2012A&A...543A.154H}
{Haberl} F.,  et~al., 2012b, \mn@doi [\aap] {10.1051/0004-6361/201218971},
  \href {http://adsabs.harvard.edu/abs/2012A%26A...543A.154H} {543, A154}

\bibitem[\protect\citeauthoryear{{Haberl} et~al.,}{{Haberl}
  et~al.}{2012c}]{2012A&A...545A.128H}
{Haberl} F.,  et~al., 2012c, \mn@doi [\aap] {10.1051/0004-6361/201219758},
  \href {http://cdsads.u-strasbg.fr/abs/2012A%26A...545A.128H} {545, A128}

\bibitem[\protect\citeauthoryear{{Han}, {Wang}, {Xu}  \& {Han}}{{Han}
  et~al.}{2016}]{2016RAA....16..159H}
{Han} J.,  {Wang} C.,  {Xu} J.,   {Han} J.-L.,  2016, \mn@doi [Research in
  Astronomy and Astrophysics] {10.1088/1674-4527/16/10/159}, \href
  {http://adsabs.harvard.edu/abs/2016RAA....16..159H} {16, 159}

\bibitem[\protect\citeauthoryear{{Hancock}, {Murphy}, {Gaensler}, {Hopkins}  \&
  {Curran}}{{Hancock} et~al.}{2012}]{2012MNRAS.422.1812H}
{Hancock} P.~J.,  {Murphy} T.,  {Gaensler} B.~M.,  {Hopkins} A.,   {Curran}
  J.~R.,  2012, \mn@doi [\mnras] {10.1111/j.1365-2966.2012.20768.x}, \href
  {http://adsabs.harvard.edu/abs/2012MNRAS.422.1812H} {422, 1812}

\bibitem[\protect\citeauthoryear{{Hancock}, {Trott}  \&
  {Hurley-Walker}}{{Hancock} et~al.}{2018}]{2018PASA...35...11H}
{Hancock} P.~J.,  {Trott} C.~M.,   {Hurley-Walker} N.,  2018, \mn@doi [\pasa]
  {10.1017/pasa.2018.3}, \href
  {http://cdsads.u-strasbg.fr/abs/2018PASA...35...11H} {35, e011}

\bibitem[\protect\citeauthoryear{{Harris} \& {Zaritsky}}{{Harris} \&
  {Zaritsky}}{2004}]{2004AJ....127.1531H}
{Harris} J.,  {Zaritsky} D.,  2004, \mn@doi [\aj] {10.1086/381953}, \href
  {http://adsabs.harvard.edu/abs/2004AJ....127.1531H} {127, 1531}

\bibitem[\protect\citeauthoryear{{Holland-Ashford}, {Lopez}, {Auchettl},
  {Temim}  \& {Ramirez-Ruiz}}{{Holland-Ashford} et~al.}{2017}]{Holland2017}
{Holland-Ashford} T.,  {Lopez} L.~A.,  {Auchettl} K.,  {Temim} T.,
  {Ramirez-Ruiz} E.,  2017, \mn@doi [\apj] {10.3847/1538-4357/aa7a5c}, \href
  {http://adsabs.harvard.edu/abs/2017ApJ...844...84H} {844, 84}

\bibitem[\protect\citeauthoryear{{Hurley-Walker} et~al.,}{{Hurley-Walker}
  et~al.}{2017}]{2017MNRAS.464.1146H}
{Hurley-Walker} N.,  et~al., 2017, \mn@doi [\mnras] {10.1093/mnras/stw2337},
  \href {http://adsabs.harvard.edu/abs/2017MNRAS.464.1146H} {464, 1146}

\bibitem[\protect\citeauthoryear{{Kargaltsev} \& {Pavlov}}{{Kargaltsev} \&
  {Pavlov}}{2008}]{Kargaltsev2008}
{Kargaltsev} O.,  {Pavlov} G.~G.,  2008, in {Bassa} C.,  {Wang} Z.,  {Cumming}
  A.,   {Kaspi} V.~M.,  eds,  American Institute of Physics Conference Series
  Vol. 983, 40 Years of Pulsars: Millisecond Pulsars, Magnetars and More. pp
  171--185 (\mn@eprint {arXiv} {0801.2602}), \mn@doi{10.1063/1.2900138}

\bibitem[\protect\citeauthoryear{{Kargaltsev}, {Rangelov}  \&
  {Pavlov}}{{Kargaltsev} et~al.}{2013}]{2013arXiv1305.2552K}
{Kargaltsev} O.,  {Rangelov} B.,   {Pavlov} G.~G.,  2013, preprint, \href
  {https://ui.adsabs.harvard.edu/#abs/2013arXiv1305.2552K} {p. arXiv:1305.2552}
  (\mn@eprint {arXiv} {1305.2552})

\bibitem[\protect\citeauthoryear{{Kargaltsev}, {Pavlov}, {Klingler}  \&
  {Rangelov}}{{Kargaltsev} et~al.}{2017}]{Kargaltsev2017}
{Kargaltsev} O.,  {Pavlov} G.~G.,  {Klingler} N.,   {Rangelov} B.,  2017,
  \mn@doi [Journal of Plasma Physics] {10.1017/S0022377817000630}, \href
  {http://adsabs.harvard.edu/abs/2017JPlPh..83e6301K} {83, 635830501}

\bibitem[\protect\citeauthoryear{{Katsuda} et~al.,}{{Katsuda}
  et~al.}{2018}]{2018ApJ...856...18K}
{Katsuda} S.,  et~al., 2018, \mn@doi [\apj] {10.3847/1538-4357/aab092}, \href
  {http://cdsads.u-strasbg.fr/abs/2018ApJ...856...18K} {856, 18}

\bibitem[\protect\citeauthoryear{{Kavanagh} et~al.,}{{Kavanagh}
  et~al.}{2016}]{2016A&A...586A...4K}
{Kavanagh} P.~J.,  et~al., 2016, \mn@doi [\aap] {10.1051/0004-6361/201527414},
  \href {http://adsabs.harvard.edu/abs/2016A%26A...586A...4K} {586, A4}

\bibitem[\protect\citeauthoryear{{Klinger}, {Dickel}, {Fields}  \&
  {Milne}}{{Klinger} et~al.}{2002}]{2002AJ....124.2135K}
{Klinger} R.~J.,  {Dickel} J.~R.,  {Fields} B.~D.,   {Milne} D.~K.,  2002,
  \mn@doi [\aj] {10.1086/341373}, \href
  {http://adsabs.harvard.edu/abs/2002AJ....124.2135K} {124, 2135}

\bibitem[\protect\citeauthoryear{{Klingler}, {Kargaltsev}, {Pavlov}, {Ng},
  {Beniamini}  \& {Volkov}}{{Klingler} et~al.}{2018}]{2018ApJ...861....5K}
{Klingler} N.,  {Kargaltsev} O.,  {Pavlov} G.~G.,  {Ng} C.-Y.,  {Beniamini} P.,
    {Volkov} I.,  2018, \mn@doi [\apj] {10.3847/1538-4357/aac6e0}, \href
  {http://adsabs.harvard.edu/abs/2018ApJ...861....5K} {861, 5}

\bibitem[\protect\citeauthoryear{{Koo} et~al.,}{{Koo}
  et~al.}{2007}]{2007PASJ...59S.455K}
{Koo} B.-C.,  et~al., 2007, \mn@doi [\pasj] {10.1093/pasj/59.sp2.S455}, \href
  {http://adsabs.harvard.edu/abs/2007PASJ...59S.455K} {59, S455}

\bibitem[\protect\citeauthoryear{Kothes}{Kothes}{2017}]{Kothes2017}
Kothes R.,  2017, Radio Properties of Pulsar Wind Nebulae.
Springer International Publishing, Cham, pp 1--27,
  \mn@doi{10.1007/978-3-319-63031-1_1}, \url
  {https://doi.org/10.1007/978-3-319-63031-1_1}

\bibitem[\protect\citeauthoryear{{Laki{\'c}evi{\'c}}
  et~al.,}{{Laki{\'c}evi{\'c}} et~al.}{2015}]{2015ApJ...799...50L}
{Laki{\'c}evi{\'c}} M.,  et~al., 2015, \mn@doi [\apj]
  {10.1088/0004-637X/799/1/50}, \href
  {http://adsabs.harvard.edu/abs/2015ApJ...799...50L} {799, 50}

\bibitem[\protect\citeauthoryear{Macri, Stanek, Bersier, Greenhill  \&
  Reid}{Macri et~al.}{2006}]{macri2006new}
Macri L.~M.,  Stanek K.,  Bersier D.,  Greenhill L.,   Reid M.,  2006, The
  Astrophysical Journal, 652, 1133

\bibitem[\protect\citeauthoryear{{Maggi} et~al.,}{{Maggi}
  et~al.}{2016}]{2016A&A...585A.162M}
{Maggi} P.,  et~al., 2016, \mn@doi [\aap] {10.1051/0004-6361/201526932}, \href
  {http://adsabs.harvard.edu/abs/2016A%26A...585A.162M} {585, A162}

\bibitem[\protect\citeauthoryear{{Maitra}, {Ballet}, {Filipovi{\'c}}, {Haberl},
  {Tiengo}, {Grieve}  \& {Roper}}{{Maitra} et~al.}{2015}]{2015A&A...584A..41M}
{Maitra} C.,  {Ballet} J.,  {Filipovi{\'c}} M.~D.,  {Haberl} F.,  {Tiengo} A.,
  {Grieve} K.,   {Roper} Q.,  2015, \mn@doi [\aap]
  {10.1051/0004-6361/201526458}, \href
  {http://adsabs.harvard.edu/abs/2015A%26A...584A..41M} {584, A41}

\bibitem[\protect\citeauthoryear{{Maitra}, {Ballet}, {Esposito}, {Haberl},
  {Tiengo}, {Filipovi{\'c}}  \& {Acero}}{{Maitra}
  et~al.}{2018}]{2018A&A...612A..87M}
{Maitra} C.,  {Ballet} J.,  {Esposito} P.,  {Haberl} F.,  {Tiengo} A.,
  {Filipovi{\'c}} M.~D.,   {Acero} F.,  2018, \mn@doi [\aap]
  {10.1051/0004-6361/201732239}, \href
  {https://ui.adsabs.harvard.edu/\#abs/2018A&A...612A..87M} {612, A87}

\bibitem[\protect\citeauthoryear{{Marelli} et~al.,}{{Marelli}
  et~al.}{2013}]{2013ApJ...765...36M}
{Marelli} M.,  et~al., 2013, \mn@doi [\apj] {10.1088/0004-637X/765/1/36}, \href
  {https://ui.adsabs.harvard.edu/#abs/2013ApJ...765...36M} {765, 36}

\bibitem[\protect\citeauthoryear{{Martin}, {Lu}, {Voelk}, {Renaud}  \&
  {Filipovic}}{{Martin} et~al.}{2019}]{cta1}
{Martin} P.,  {Lu} C.-C.,  {Voelk} H.,  {Renaud} M.,   {Filipovic} M.,  2019,
  KSP: Large Magellanic Cloud Survey.
pp 125--141 (\mn@eprint {}
  {https://www.worldscientific.com/doi/pdf/10.1142/97898132700910007}),
  \mn@doi{10.1142/9789813270091_0007}, \url
  {https://www.worldscientific.com/doi/abs/10.1142/97898132700910007}

\bibitem[\protect\citeauthoryear{{Mattana} et~al.,}{{Mattana}
  et~al.}{2009}]{2009ApJ...694...12M}
{Mattana} F.,  et~al., 2009, \mn@doi [\apj] {10.1088/0004-637X/694/1/12}, \href
  {https://ui.adsabs.harvard.edu/#abs/2009ApJ...694...12M} {694, 12}

\bibitem[\protect\citeauthoryear{{McClure-Griffiths}
  et~al.,}{{McClure-Griffiths} et~al.}{2018}]{2018NatAs...2..901M}
{McClure-Griffiths} N.~M.,  et~al., 2018, \mn@doi [Nature Astronomy]
  {10.1038/s41550-018-0608-8}, \href
  {https://ui.adsabs.harvard.edu/#abs/2018NatAs...2..901M} {2, 901}

\bibitem[\protect\citeauthoryear{{McKee} \& {Truelove}}{{McKee} \&
  {Truelove}}{1995}]{1995PhR...256..157M}
{McKee} C.~F.,  {Truelove} J.~K.,  1995, \mn@doi [\physrep]
  {10.1016/0370-1573(94)00106-D}, \href
  {https://ui.adsabs.harvard.edu/#abs/1995PhR...256..157M} {256, 157}

\bibitem[\protect\citeauthoryear{{Moon} et~al.,}{{Moon}
  et~al.}{2004}]{2004ApJ...610L..33M}
{Moon} D.~S.,  et~al., 2004, \mn@doi [\apj] {10.1086/423238}, \href
  {https://ui.adsabs.harvard.edu/\#abs/2004ApJ...610L..33M} {610, L33}

\bibitem[\protect\citeauthoryear{{Ng} \& {Romani}}{{Ng} \&
  {Romani}}{2007}]{Ng2007}
{Ng} C.-Y.,  {Romani} R.~W.,  2007, \mn@doi [\apj] {10.1086/513597}, \href
  {http://adsabs.harvard.edu/abs/2007ApJ...660.1357N} {660, 1357}

\bibitem[\protect\citeauthoryear{{Owen} et~al.,}{{Owen}
  et~al.}{2011}]{2011A&A...530A.132O}
{Owen} R.~A.,  et~al., 2011, \mn@doi [\aap] {10.1051/0004-6361/201116586},
  \href {http://adsabs.harvard.edu/abs/2011A%26A...530A.132O} {530, A132}

\bibitem[\protect\citeauthoryear{{Pannuti} et~al.,}{{Pannuti}
  et~al.}{2017}]{2017ApJ...839...59P}
{Pannuti} T.~G.,  et~al., 2017, \mn@doi [\apj] {10.3847/1538-4357/aa615c},
  \href {https://ui.adsabs.harvard.edu/\#abs/2017ApJ...839...59P} {839, 59}

\bibitem[\protect\citeauthoryear{{Pavan} et~al.,}{{Pavan}
  et~al.}{2014a}]{2014IJMPS..2860172P}
{Pavan} L.,  et~al., 2014a, in International Journal of Modern Physics
  Conference Series. p. 1460172, \mn@doi{10.1142/S2010194514601720}

\bibitem[\protect\citeauthoryear{{Pavan} et~al.,}{{Pavan}
  et~al.}{2014b}]{2014A&A...562A.122P}
{Pavan} L.,  et~al., 2014b, \mn@doi [\aap] {10.1051/0004-6361/201322588}, \href
  {http://adsabs.harvard.edu/abs/2014A%26A...562A.122P} {562, A122}

\bibitem[\protect\citeauthoryear{{Payne}, {Filipovi{\'c}}, {Reid}, {Jones},
  {Staveley-Smith}  \& {White}}{{Payne} et~al.}{2004}]{2004MNRAS.355...44P}
{Payne} J.~L.,  {Filipovi{\'c}} M.~D.,  {Reid} W.,  {Jones} P.~A.,
  {Staveley-Smith} L.,   {White} G.~L.,  2004, \mn@doi [\mnras]
  {10.1111/j.1365-2966.2004.08287.x}, \href
  {http://adsabs.harvard.edu/abs/2004MNRAS.355...44P} {355, 44}

\bibitem[\protect\citeauthoryear{{Payne}, {White}, {Filipovi{\'c}}  \&
  {Pannuti}}{{Payne} et~al.}{2007}]{2007MNRAS.376.1793P}
{Payne} J.~L.,  {White} G.~L.,  {Filipovi{\'c}} M.~D.,   {Pannuti} T.~G.,
  2007, \mn@doi [\mnras] {10.1111/j.1365-2966.2007.11561.x}, \href
  {http://adsabs.harvard.edu/abs/2007MNRAS.376.1793P} {376, 1793}

\bibitem[\protect\citeauthoryear{{Pellegrini}, {Oey}, {Winkler}, {Points},
  {Smith}, {Jaskot}  \& {Zastrow}}{{Pellegrini} et~al.}{2012}]{Pellegrini2012}
{Pellegrini} E.~W.,  {Oey} M.~S.,  {Winkler} P.~F.,  {Points} S.~D.,  {Smith}
  R.~C.,  {Jaskot} A.~E.,   {Zastrow} J.,  2012, \mn@doi [\apj]
  {10.1088/0004-637X/755/1/40}, \href
  {http://adsabs.harvard.edu/abs/2012ApJ...755...40P} {755, 40}

\bibitem[\protect\citeauthoryear{{Ransom}}{{Ransom}}{2001}]{2001AAS...19911903R}
{Ransom} S.~M.,  2001, in American Astronomical Society Meeting Abstracts. p.
  119.03

\bibitem[\protect\citeauthoryear{{Reach} et~al.,}{{Reach}
  et~al.}{2006}]{2006AJ....131.1479R}
{Reach} W.~T.,  et~al., 2006, \mn@doi [\aj] {10.1086/499306}, \href
  {http://adsabs.harvard.edu/abs/2006AJ....131.1479R} {131, 1479}

\bibitem[\protect\citeauthoryear{{Reid}, {Payne}, {Filipovi{\'c}}, {Danforth},
  {Jones}, {White}  \& {Staveley-Smith}}{{Reid}
  et~al.}{2006}]{2006MNRAS.367.1379R}
{Reid} W.~A.,  {Payne} J.~L.,  {Filipovi{\'c}} M.~D.,  {Danforth} C.~W.,
  {Jones} P.~A.,  {White} G.~L.,   {Staveley-Smith} L.,  2006, \mn@doi [\mnras]
  {10.1111/j.1365-2966.2006.10017.x}, \href
  {http://adsabs.harvard.edu/abs/2006MNRAS.367.1379R} {367, 1379}

\bibitem[\protect\citeauthoryear{{Reynolds}, {Pavlov}, {Kargaltsev},
  {Klingler}, {Renaud}  \& {Mereghetti}}{{Reynolds}
  et~al.}{2017}]{2017SSRv..207..175R}
{Reynolds} S.~P.,  {Pavlov} G.~G.,  {Kargaltsev} O.,  {Klingler} N.,  {Renaud}
  M.,   {Mereghetti} S.,  2017, \mn@doi [\ssr] {10.1007/s11214-017-0356-6},
  \href {http://adsabs.harvard.edu/abs/2017SSRv..207..175R} {207, 175}

\bibitem[\protect\citeauthoryear{{Rho} \& {Petre}}{{Rho} \&
  {Petre}}{1998}]{1998ApJ...503L.167R}
{Rho} J.,  {Petre} R.,  1998, \mn@doi [\apj] {10.1086/311538}, \href
  {https://ui.adsabs.harvard.edu/\#abs/1998ApJ...503L.167R} {503, L167}

\bibitem[\protect\citeauthoryear{{Roper}, {McEntaffer}, {DeRoo}, {Filipovic},
  {Wong}  \& {Crawford}}{{Roper} et~al.}{2015}]{2015ApJ...803..106R}
{Roper} Q.,  {McEntaffer} R.~L.,  {DeRoo} C.,  {Filipovic} M.,  {Wong} G.~F.,
  {Crawford} E.~J.,  2015, \mn@doi [\apj] {10.1088/0004-637X/803/2/106}, \href
  {http://adsabs.harvard.edu/abs/2015ApJ...803..106R} {803, 106}

\bibitem[\protect\citeauthoryear{{Russell} \& {Dopita}}{{Russell} \&
  {Dopita}}{1992}]{1992ApJ...384..508R}
{Russell} S.~C.,  {Dopita} M.~A.,  1992, \mn@doi [\apj] {10.1086/170893}, \href
  {http://adsabs.harvard.edu/abs/1992ApJ...384..508R} {384, 508}

\bibitem[\protect\citeauthoryear{{Safi-Harb}, {Ogelman}  \&
  {Finley}}{{Safi-Harb} et~al.}{1995}]{1995ApJ...439..722S}
{Safi-Harb} S.,  {Ogelman} H.,   {Finley} J.~P.,  1995, \mn@doi [\apj]
  {10.1086/175212}, \href
  {https://ui.adsabs.harvard.edu/\#abs/1995ApJ...439..722S} {439, 722}

\bibitem[\protect\citeauthoryear{{Sault} \& {Wieringa}}{{Sault} \&
  {Wieringa}}{1994}]{1994A&AS..108..585S}
{Sault} R.~J.,  {Wieringa} M.~H.,  1994, \aaps, \href
  {http://adsabs.harvard.edu/abs/1994A%26AS..108..585S} {108, 585}

\bibitem[\protect\citeauthoryear{{Sault}, {Teuben}  \& {Wright}}{{Sault}
  et~al.}{1995}]{1995ASPC...77..433S}
{Sault} R.~J.,  {Teuben} P.~J.,   {Wright} M.~C.~H.,  1995, in {Shaw} R.~A.,
  {Payne} H.~E.,   {Hayes} J.~J.~E.,  eds,  Astronomical Society of the Pacific
  Conference Series Vol. 77, Astronomical Data Analysis Software and Systems
  IV. p.~433 (\mn@eprint {} {astro-ph/0612759})

\bibitem[\protect\citeauthoryear{Scowcroft, Freedman, Madore, Monson, Persson,
  Rich, Seibert  \& Rigby}{Scowcroft et~al.}{2016}]{0004-637X-816-2-49}
Scowcroft V.,  Freedman W.~L.,  Madore B.~F.,  Monson A.,  Persson S.~E.,  Rich
  J.,  Seibert M.,   Rigby J.~R.,  2016, The Astrophysical Journal, 816, 49

\bibitem[\protect\citeauthoryear{{Smith}, {Leiton}  \& {Pizarro}}{{Smith}
  et~al.}{2000}]{2000ASPC..221...83S}
{Smith} C.,  {Leiton} R.,   {Pizarro} S.,  2000, in {Alloin} D.,  {Olsen} K.,
  {Galaz} G.,  eds,  Astronomical Society of the Pacific Conference Series Vol.
  221, Stars, Gas and Dust in Galaxies: Exploring the Links. p.~83

\bibitem[\protect\citeauthoryear{{Stanimirovic}, {Staveley-Smith}, {Dickey},
  {Sault}  \& {Snowden}}{{Stanimirovic} et~al.}{1999}]{Stanimiro1999}
{Stanimirovic} S.,  {Staveley-Smith} L.,  {Dickey} J.~M.,  {Sault} R.~J.,
  {Snowden} S.~L.,  1999, \mn@doi [\mnras] {10.1046/j.1365-8711.1999.02013.x},
  \href {http://adsabs.harvard.edu/abs/1999MNRAS.302..417S} {302, 417}

\bibitem[\protect\citeauthoryear{{Staveley-Smith}, {Sault}, {Hatzidimitriou},
  {Kesteven}  \& {McConnell}}{{Staveley-Smith}
  et~al.}{1997}]{1997MNRAS.289..225S}
{Staveley-Smith} L.,  {Sault} R.~J.,  {Hatzidimitriou} D.,  {Kesteven} M.~J.,
  {McConnell} D.,  1997, \mn@doi [\mnras] {10.1093/mnras/289.2.225}, \href
  {https://ui.adsabs.harvard.edu/#abs/1997MNRAS.289..225S} {289, 225}

\bibitem[\protect\citeauthoryear{{Str{\"u}der} et~al.,}{{Str{\"u}der}
  et~al.}{2001}]{2001A&A...365L..18S}
{Str{\"u}der} L.,  et~al., 2001, \mn@doi [\aap] {10.1051/0004-6361:20000066},
  \href {http://adsabs.harvard.edu/abs/2001A%26A...365L..18S} {365, L18}

\bibitem[\protect\citeauthoryear{{Stupar}, {Parker}, {Filipovi{\'c}}, {Frew},
  {Boji{\v c}i{\'c}}  \& {Aschenbach}}{{Stupar}
  et~al.}{2007}]{2007MNRAS.381..377S}
{Stupar} M.,  {Parker} Q.~A.,  {Filipovi{\'c}} M.~D.,  {Frew} D.~J.,  {Boji{\v
  c}i{\'c}} I.,   {Aschenbach} B.,  2007, \mn@doi [\mnras]
  {10.1111/j.1365-2966.2007.12296.x}, \href
  {http://adsabs.harvard.edu/abs/2007MNRAS.381..377S} {381, 377}

\bibitem[\protect\citeauthoryear{{Sturm} et~al.,}{{Sturm}
  et~al.}{2013}]{2013A&A...558A...3S}
{Sturm} R.,  et~al., 2013, \mn@doi [\aap] {10.1051/0004-6361/201219935}, \href
  {http://adsabs.harvard.edu/abs/2013A%26A...558A...3S} {558, A3}

\bibitem[\protect\citeauthoryear{{Temim}, {Slane}, {Reynolds}, {Raymond}  \&
  {Borkowski}}{{Temim} et~al.}{2010}]{2010ApJ...710..309T}
{Temim} T.,  {Slane} P.,  {Reynolds} S.~P.,  {Raymond} J.~C.,   {Borkowski}
  K.~J.,  2010, \mn@doi [\apj] {10.1088/0004-637X/710/1/309}, \href
  {https://ui.adsabs.harvard.edu/\#abs/2010ApJ...710..309T} {710, 309}

\bibitem[\protect\citeauthoryear{{Tiengo}, {Esposito}  \&
  {Mereghetti}}{{Tiengo} et~al.}{2008}]{Tiengo2008}
{Tiengo} A.,  {Esposito} P.,   {Mereghetti} S.,  2008, \mn@doi [\apjl]
  {10.1086/590078}, \href {http://adsabs.harvard.edu/abs/2008ApJ...680L.133T}
  {680, L133}

\bibitem[\protect\citeauthoryear{{Turner} \& {Pounds}}{{Turner} \&
  {Pounds}}{1989}]{Turner1989}
{Turner} T.~J.,  {Pounds} K.~A.,  1989, \mn@doi [\mnras]
  {10.1093/mnras/240.4.833}, \href
  {http://adsabs.harvard.edu/abs/1989MNRAS.240..833T} {240, 833}

\bibitem[\protect\citeauthoryear{{Turner} et~al.,}{{Turner}
  et~al.}{2001}]{2001A&A...365L..27T}
{Turner} M.~J.~L.,  et~al., 2001, \mn@doi [\aap] {10.1051/0004-6361:20000087},
  \href {http://adsabs.harvard.edu/abs/2001A%26A...365L..27T} {365, L27}

\bibitem[\protect\citeauthoryear{{Turtle}, {Ye}, {Amy}  \& {Nicholls}}{{Turtle}
  et~al.}{1998}]{1998PASA...15..280T}
{Turtle} A.~J.,  {Ye} T.,  {Amy} S.~W.,   {Nicholls} J.,  1998, \mn@doi [\pasa]
  {10.1071/AS98280}, \href {http://adsabs.harvard.edu/abs/1998PASA...15..280T}
  {15, 280}

\bibitem[\protect\citeauthoryear{{Weaver}, {McCray}, {Castor}, {Shapiro}  \&
  {Moore}}{{Weaver} et~al.}{1977}]{1977ApJ...218..377W}
{Weaver} R.,  {McCray} R.,  {Castor} J.,  {Shapiro} P.,   {Moore} R.,  1977,
  \mn@doi [\apj] {10.1086/155692}, \href
  {http://adsabs.harvard.edu/abs/1977ApJ...218..377W} {218, 377}

\bibitem[\protect\citeauthoryear{{Williams} et~al.,}{{Williams}
  et~al.}{2008}]{2008ApJ...687.1054W}
{Williams} B.~J.,  et~al., 2008, \mn@doi [\apj] {10.1086/592139}, \href
  {https://ui.adsabs.harvard.edu/\#abs/2008ApJ...687.1054W} {687, 1054}

\bibitem[\protect\citeauthoryear{{Wong}, {Filipovic}, {Crawford}, {de Horta},
  {Galvin}, {Draskovic}  \& {Payne}}{{Wong}
  et~al.}{2011a}]{2011SerAJ.182...43W}
{Wong} G.~F.,  {Filipovic} M.~D.,  {Crawford} E.~J.,  {de Horta} A.~Y.,
  {Galvin} T.,  {Draskovic} D.,   {Payne} J.~L.,  2011a, \mn@doi [Serbian
  Astronomical Journal] {10.2298/SAJ1182043W}, \href
  {http://adsabs.harvard.edu/abs/2011SerAJ.182...43W} {182, 43}

\bibitem[\protect\citeauthoryear{{Wong} et~al.,}{{Wong}
  et~al.}{2011b}]{2011SerAJ.183..103W}
{Wong} G.~F.,  et~al., 2011b, \mn@doi [Serbian Astronomical Journal]
  {10.2298/SAJ1183103W}, \href
  {http://adsabs.harvard.edu/abs/2011SerAJ.183..103W} {183, 103}

\bibitem[\protect\citeauthoryear{{Wong} et~al.,}{{Wong}
  et~al.}{2012a}]{2012SerAJ.184...93W}
{Wong} G.~F.,  et~al., 2012a, \mn@doi [Serbian Astronomical Journal]
  {10.2298/SAJ1284093W}, \href
  {http://adsabs.harvard.edu/abs/2012SerAJ.184...93W} {184, 93}

\bibitem[\protect\citeauthoryear{{Wong}, {Filipovic}, {Crawford}, {Tothill},
  {De Horta}  \& {Galvin}}{{Wong} et~al.}{2012b}]{2012SerAJ.185...53W}
{Wong} G.~F.,  {Filipovic} M.~D.,  {Crawford} E.~J.,  {Tothill} N.~F.~H.,  {De
  Horta} A.~Y.,   {Galvin} T.~J.,  2012b, \mn@doi [Serbian Astronomical
  Journal] {10.2298/SAJ1285053W}, \href
  {http://adsabs.harvard.edu/abs/2012SerAJ.185...53W} {185, 53}

\makeatother
\end{thebibliography}
\end{document}